# The Multiview Observatory for Solar Terrestrial Science (MOST)

N. Gopalswamy[1*], S. Christe[1], S. F. Fung[1], Q. Gong[1], J. R. Gruesbeck[1], L. K. Jian[1], S. G. Kanekal[1], C. Kay[1,2], T. A. Kucera[1], J. E. Leake[1], L. Li[1], P. Mäkelä[1,2], P. Nikulla[1], N. L. Reginald[1,2], A. Shih[1], S. K. Tadikonda[1], N. Viall[1], L. B. Wilson III[1], S. Yashiro[1,2], L. Golub[3], E. DeLuca[3], K. Reeves[3], A. C. Sterling[4], A. R. Winebarger[4], C. DeForest[5], D. M. Hassler[5], D. B. Seaton[5], M. I. Desai[6], P. S. Mokashi[6], J. Lazio[7], E. A. Jensen[8], W. B. Manchester[9], N. Sachdeva[9], B. Wood[10], J. Kooi[10], P. Hess[10], D. B. Wexler[11], S. D. Bale[12], S. Krucker[12], N. Hurlburt[13], M. DeRosa[13], S. Gosain[14], K. Jain[14], S. Kholikov[14], G. J. D. Petrie[14], A. Pevtsov[14], S. C. Tripathy[14], J. Zhao[15], P. H. Scherrer[15], S. P. Rajaguru[15], T. Woods[16], M. Kenney[16], J. Zhang[17], C. Scolini[18], K.-S. Cho[19], Y.-D. Park[19], B. V. Jackson[20]

[1]NASA Goddard Space Flight Center, Greenbelt, MD, United States

[2]The Catholic University of America, Washington, DC, United States

[3]Center for Astrophysics | Harvard &Smithsonian, Cambridge, MA, United States

[4]NASA Marshall Space Flight Center, Huntsville, AL, United States

[5]Southwest Research Institute, Boulder, CO, United States

[6]Southwest Research Institute, San Antonio, TX, United States

[7]Jet Propulsion Laboratory, Pasadena, CA, United States

[8]Planetary Science Institute, Tucson, AZ, United States

[9]University of Michigan, Ann Arbor, MI, United States

[10]Naval Research Laboratory, Washington, DC, United States

[11]University of Massachusetts Lowell, Lowell, MA, United States

[12]University of California, Berkeley, CA, United States

[13]Lockheed Martin Advanced Technology Center, Palo Alto, CA, United States

[14]National Solar Observatory, Boulder, CO, United States

[15]Stanford University, Stanford, CA, United States

[16]University of Colorado, Boulder, CO, United States

[17]George Mason University, Fair Fax, VA, United States

[18]University of New Hampshire, Durham, NH, United States

[19]Korea Astronomy and Space Science Institute, Daejeon, Republic of Korea

[20]University of California San Diego, La Jolla, CA, United States

\* **Correspondence:**
Nat Gopalswamy
nat.gopalswamy@nasa.gov

**Abstract**

We report on a study of the Multiview Observatory for Solar Terrestrial Science (MOST) mission that will provide comprehensive imagery and time series data needed to understand the magnetic connection between the solar interior and the solar atmosphere/inner heliosphere. MOST will build upon the successes of SOHO and STEREO missions with new views of the Sun and enhanced instrument capabilities. This article is based on a study conducted at NASA Goddard Space Flight Center that determined the required instrument refinement, spacecraft accommodation, launch configuration, and flight dynamics for mission success. MOST is envisioned as the next generation great observatory positioned to obtain three-dimensional information of large-scale heliospheric structures such as coronal mass ejections, stream interaction regions, and the solar wind itself. The MOST mission consists of 2 pairs of spacecraft located in the vicinity of Sun-Earth Lagrange points L4 (MOST1, MOST3) and L5 (MOST2 and MOST4). The spacecraft stationed at L4 (MOST1) and L5 (MOST2) will each carry seven remote-sensing and three in-situ instrument suites, including a novel radio package known as the Faraday Effect Tracker of Coronal and Heliospheric structures (FETCH). MOST3 and MOST4 will carry only the FETCH instruments and are positioned at variable locations along the Earth orbit up to 20° ahead of L4 and 20° behind L5, respectively. FETCH will have polarized radio transmitters and receivers on all four spacecraft to measure the magnetic content of solar wind structures propagating from the Sun to Earth using the Faraday rotation technique. The MOST mission will be able to sample the magnetized plasma throughout the Sun-Earth connected space during the mission lifetime over a solar cycle.

## 1    Introduction

The Sun is an ordinary star, but it is unique and vital to life on Earth. The magnetic variability of the Sun affects human technology in space and on ground. The Sun is the only star that can be observed in detail through both remote-sensing and in-situ techniques and hence contributes toward the understanding of stellar phenomena. Unprecedented advances in heliophysics made possible by great observatories such as the Solar and Heliospheric Observatory (SOHO, Domingo et al. 1995), Solar Terrestrial Relations Observatory (STEREO, Kaiser et al. 2008), and Solar Dynamics Observatory (SDO, Pesnell et al. 2012) have demonstrated the need for comprehensive observations that can enable the science of a large swath of the community. These observatories helped us accumulate a wealth of knowledge on solar and heliospheric structures. However, many fundamental questions remain unanswered: What are the changes that occur in the convection zone before active regions emerge? Why does flux emerge in a large-scale forming active regions? How do magnetic fields become energized to erupt and what processes initiate the eruptions? How do solar eruptions result in particle acceleration, alone and in combination with flare reconnection? How does shock geometry and magnitude evolve and how does this relate to solar energetic particles (SEPs) and radio bursts? What is the radial density profile of shock-driving coronal mass ejections (CMEs) from the nose to trailing edge? How do CMEs and corotating interaction regions (CIRs) evolve in the inner heliosphere? What are the implications of the interchange reconnection taking place between open and closed field lines? What is the internal magnetic structure of CMEs that cause geomagnetic storms at Earth? Clearly, many of these questions involve solar magnetic fields at various layers of the solar atmosphere and we do not have sufficient knowledge about them.

Simulations show a dramatic improvement in accurately capturing solar wind structure when provided with improved magnetic observations including observational coverage of the poles (Petrie et al., 2018; Pevtsov et al., 2020). Clear improvements are already achieved when the Sun can be observed from Sun-Earth Lagrange points L1, L4, and L5 providing coverage of over 65% of the solar surface. Wider and longer duration Doppler coverage of the solar surface from L1, L4 and L5 views will provide the necessary signal-to-noise for helioseismic localization of non-axisymmetric changes in flow patterns in the convection zone. In the photosphere, plasma controls the magnetic field, while the control switches to the magnetic field in the chromosphere. Thus, extending the magnetic field measurements to the chromosphere provides information on the magnetic roots of large-scale coronal structures and adds fidelity to coronal/heliospheric models. Currently, we obtain the magnetic flux over only a 60° to 90° width wedge in longitude centered on the Sun-Earth line, while what is ideally needed is over the entire solar surface. While coronal magnetic field measurement techniques are maturing but largely lacking, substantial progress can still be made with routine photospheric and chromospheric magnetic field measurements. Far away from the Sun, magnetic fields are measured in situ by spacecraft at L1. Parker Solar Probe (Fox et al. 2016) and Solar Orbiter (Müller et al. 2020) provide information at several locations in the inner heliosphere, but not systematically. Faraday rotation (FR) provides a different and unique way to measure magnetic field in large-scale coronal and heliospheric structures by transmitting and receiving spacecraft radio signals through such structures. By suitable frequency and antenna choices, one can probe structures over the Sun-Earth distance. This paper outlines the concept of a mission called the Multiview Observatory for Solar Terrestrial Science (MOST) that will provide comprehensive imagery and time series data needed to understand the magnetic connection between the solar interior and the atmosphere. MOST will build upon the successes of SOHO and STEREO with new views from L4 and L5 and from the vicinity of those points.

In this paper, we present the results of a mission study undertaken at NASA's Goddard Space Flight Center (GSFC) that focused on optimized science payload, instrument accommodation, flight dynamics, and launch system. This paper is organized as follows. The science objectives and a mission overview are presented in section 2. The scientific payload, its accommodation on the spacecraft, and flight dynamics are described in section 3. Section 4 describes synergy among instruments and modeling, while a brief description of the mission operations is given in section 5. The summary and conclusions are presented in section 6.

## 2  Scientific Objectives and the MOST Mission Overview

In this section, we identify the science questions and develop objectives that need to be achieved to answer these questions. To achieve the objectives, we identify the instrument and mission requirements. These tasks are performed by identifying gaps in the past measurements and characterize the optimal set of instruments. We develop high level design of the required instruments improving on the past designs and employing new technologies that have become available in the recent past. We design the spacecraft to accommodate the instruments, the fairing to accommodate the spacecraft, and the launch vehicle. We perform flight dynamics analysis and select the orbit for the mission. Finally we estimate the cost of the mission based on a previous study.

## 2.1 Goals, Objectives, and the Science Traceability Matrix

The MOST mission concept draws heavily on the success of great observatories such as SOHO, STEREO, and SDO and combines the capabilities to build the next generation great observatory. These observatories have demonstrated the value of sustained observations that have greatly added to our knowledge of the variable solar-terrestrial system (see e.g., Temmer 2021). This advance can be accelerated by devising a new mission that implements additional capabilities that were not included in SOHO and STEREO. Since the primary cause of variability in the solar-heliospheric system is solar magnetism, measuring the magnetic field at the Sun and in the surrounding heliosphere is of utmost importance. Therefore, the primary science goal of MOST is to understand the magnetic coupling of the solar interior to the heliosphere. As noted in the introduction, there are many unanswered fundamental questions that form the basis for formulating the science objectives of the MOST mission. The fundamental questions can be grouped into a set of three high-level science questions related to solar and heliospheric magnetic fields, solar eruptions, and the solar wind. The mission objectives and the underlying science questions are listed in the MOST Science Traceability Matrix (STM, see Table 1).

**Table 1.** MOST Science Traceability Matrix.

| Science Question | Objectives | Measurement Requirements | Instrument Requirements | Mission Requirements |
|---|---|---|---|---|
| 1. How do active regions evolve before and after emerging to the solar surface? | 1.1 Derive the physical properties of the convection zone helioseismically | Dopplergrams (velocities better than 20 m/s in each 0.5 $Mm^2$ pixel) from viewing angles separated by 30°-90° | Sun-pointed telescope to obtain full disk images with 1" pixels; 1-min cadence | Identical telescopes on MOST1&2. Telescope at Earth/L1 assumed |
| | 1.2 Determine the complete life cycle of active regions | Photospheric and chromospheric line-of-sight (LOS) magnetograms from viewing angles separated by 30°-90° | Sun-pointed telescope to obtain full disk images with 1" pixels; 1-min cadence | Identical telescopes on MOST1&2. Telescope at Earth/L1 assumed |
| | 1.3 Determine the global magnetic field distribution on the Sun | LOS magnetograms from viewing angles separated by 30°-90° angles to cover at least | Sun-pointed telescope to obtain full disk images with 1" pixels; 90-min cadence | Identical telescopes on MOST1&2. Telescope at Earth/L1 assumed |
| | | 65% of the solar surface | | |
| 2. How do CME flux ropes form, accelerate, drive shocks, and evolve from near the Sun into the heliosphere including particle acceleration? | 2.1 Track and characterize 3-D CME acceleration and the evolution of the CME-shock complex through the outer corona and young solar wind; determine forces acting on CMEs. | EUV images at multiple wavelengths (17.1-20.5 nm); white-light (WL) coronagraph and heliospheric images from spatially separated viewing angles; in-situ plasma and B measurements; remote and local radio waves; FR angle of spacecraft (S/C) signals from multiple views (uncertainty better than ±8°) | Sun-pointed EUV imager (2" pixels; field of view (FOV) 0-3 Rs), hard X-ray imager (HXI) (3.5"-90" pixels, FOV 2°×2°) and WL coronagraph (15" pixels; FOV 2-15 Rs); Heliospheric imager (2' pixels; 3°-65°; 10 min cadence); radio telescope (0.02 to 20 MHz); radio transceiver (f in 100-200 MHz range); magnetometer; plasma analyzer | Identical set of instruments on MOST1&2. FR package with elements on MOST1&2; additional elements on MOST3&4. |
| | 2.2 Reconstruct and track flux ropes and flare structure from pre-eruption, eruption, and post-eruption data | Multiview LOS magnetogram, EUV, HXI, WL coronagraph, HI, and FR properties of S/C signals | Same as above | Same as above |
| 3. How do CIR magnetic fields evolve in the inner heliosphere and accelerate particles? | 3.1 Track longitudinal evolution of CIRs from L5 to Earth to L4 | In-situ plasma, magnetic field (B), and energetic particle measurements at L4 and L5; multiview heliospheric images and FR angle of S/C signals along multiple ray paths | Heliospheric imager (2' pixels; 3° - 65°; 10 min cadence); radio transceiver (f in 100-200 MHz range); magnetometer; plasma analyzer; particle detector | Heliospheric imagers on MOST1&2. FR package with elements on MOST1&2; additional elements on MOST3&4 |
| | 3.2 Determine role of interchange reconnection | LOS magnetogram for active region B, EUV coronal hole | Sun-pointed magnetograph (1" pixels; 1-min cadence) and | Magnetograph, EUV, Heliospheric imager, and |

| | between active region and coronal hole in providing seed particles to CIR accelerator | properties, HI, Faraday rotation angle of S/C signals, remote and local radio waves; energetic particle detector | EUV imager (1" pixels; FOV 0-3 Rs); Heliospheric imager; radio telescope; radio transceiver | EPD on MOST1&2.  FR package on MOST1, 2, 3 & 4 |
|---|---|---|---|---|

Each question in the STM (column 1) can be answered by achieving a set of science objectives listed in column 2. The measurement requirements towards achieving the objectives are listed in column 3 including the nature of the sensor to be used. The requirements on scientific instruments that make the necessary measurements are listed in column 4. Finally, column 5 sets the mission requirements.

## 2.2   Mission Overview

MOST will be a 4-spacecraft mission with one each at L4 (MOST1) and L5 (MOST2) and the other two (MOST3 and MOST4) at variable locations along Earth orbit (see Figure 1). MOST1 and MOST2 will each carry seven remote-sensing and 3 in-situ instruments. All four spacecraft will carry a novel radio package known as the Faraday Effect Tracker of Coronal and Heliospheric structures (FETCH) that will systematically probe the magnetic content of transient interplanetary structures including coronal mass ejections (CMEs) and stream interaction regions (SIRs). The Faraday rotation measurements will provide magnetic content of these structures at various heliocentric distances from the outer corona to Earth's vicinity. Photospheric and/or chromospheric magnetograms will cover >70% of the solar surface providing synchronic maps needed for accurately modeling the corona and solar wind. EUV, coronagraph, radio spectrograph, and heliospheric imager observations from multiple viewpoints provide 3-D information on CMEs/CME-driven shocks, SIRs, and other solar wind structures. Hard X-ray imagers will provide the flare aspects of solar eruptions to complement the CME aspects. In-situ instruments provide ground truth to the remote-sensing observations.  MOST will generate the following science data products: magnetograms, Dopplergrams, EUV images, hard X-ray **spectra and** images, coronagraph images, heliospheric images, radio dynamic spectra and time series, Faraday rotation time series, time series of solar wind plasma parameters, solar wind magnetic field vectors, and solar energetic particle intensity and spectra. **We believe that these data products will greatly help in tracking** the flow of energy from the Sun into the heliosphere and various physical processes that result from the energy flow.

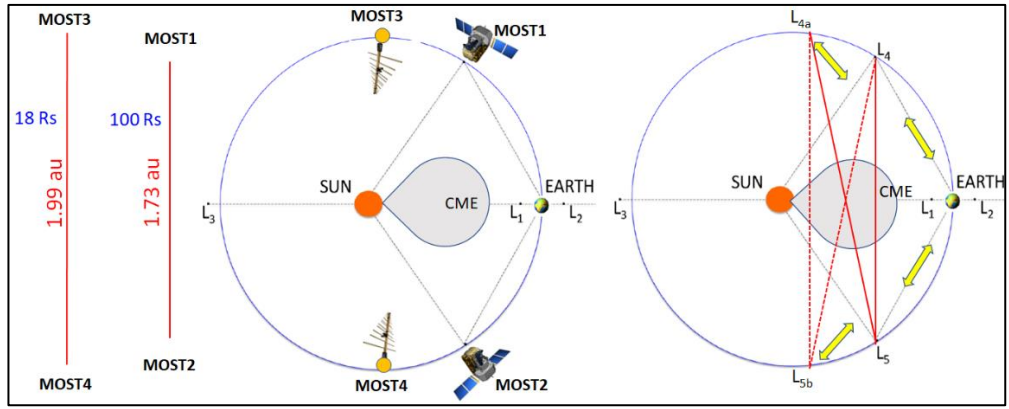

**Figure 1**. Overview of the MOST mission with the four constituent spacecraft at L4 (MOST1), L5 (MOST2), ahead of L4 (L4', MOST3) and behind L5 (L5', MOST4). MOST1&2 will have identical remote-sensing and in-situ instrument suites. MOST3&4 will carry only radio equipment for Faraday rotation measurements. The approximate MOST1-MOST2 and MOST3-MOST4 distances are shown at the left indicating the long signal paths for spacecraft radio signals (red numbers) and their closest approach to the Sun (blue numbers). The red lines in the right indicate FETCH signal paths. The yellow double arrows indicate communication links. The five Lagrange points (L1 – L5) are shown for reference.

MOST, a large 10-year mission, is well aligned with NASA's Heliophysics objectives and will provide an unprecedented opportunity to achieve the scientific objectives with broad participation from the heliophysics community. The MOST mission as described here assumes that imagery and time-series data will be available from the Sun Earth line (ground-based observatories and space-based observatories at L1). If not available, a spacecraft similar to MOST1 or MOST2 can optionally be deployed at Sun-Earth L1.

## 3  Science Instruments, their Accommodation, and Flight Dynamics

In this section we describe the optimal set of instruments, their placements in the spacecraft, launch configuration and vehicle, flights dynamics, life cycle of the mission, and costs.

### 3.1  The Science Payload

The seven remote-sensing and three in-situ instruments to be carried by each of MOST1&2 are listed in Table 2. In each case, improvements over previously flown instruments are noted as "New" in column 1. The instruments are optimized to obtain maximum information on the Sun-heliospheric system in accordance with the STM given in Table 1. The instrument suites provide imagery and time series data to reveal magnetic connectivity across solar and heliospheric domains. Actively probed Faraday rotation studies form a hybrid between in-situ methods, which provide detailed field information at each sampled point, and imaging methods, which provide mostly distributions of material density across space. Data from a combination of MOST instruments are needed for investigations that lead to achieving the science objectives. We note that most of the instruments trace their heritage to SOHO and STEREO. These instruments will be refined and improved by incorporating new developments in sensor technology. There are new instruments such as the Magnetic and Doppler Imager (MaDI) that was not included in STEREO. The Hard X-ray Imager (HXI) and the FETCH instrument are the other remote-sensing instruments not included in SOHO or STEREO instrument suites.

**Table 2**. Science instruments and their purpose

| Instrument, Heritage[a], and Improvements | Purpose |
|---|---|
| **Magnetic and Doppler Imagers (MaDI)** *SOHO, Solar Orbiter, SDO* New: Routine chromospheric magnetograms | To study surface (photosphere, chromosphere) and subsurface magnetism by combining magnetic and Doppler measurements. Also routinely obtain chromospheric magnetograms |
| **Inner Coronal Imager in EUV (ICIE)** *SWAP, SUVI* New: Extended FOV to have significant overlap with coronagraph FOV | To study active regions, coronal holes, post-eruption arcades (PEAs), coronal waves, and coronal dimming by capturing the magnetic connection between the photosphere and the corona |
| **Hard X-ray Imager (HXI)** *Solar Orbiter* | To image thermal and non-thermal component of flares and study the relationship with radio bursts and CME flux ropes |
| **White-light Coronagraph (WCOR)** *STEREO, BITSE* New: Polarization detector, two-stage optics | To track quiescent and transient coronal structures seamlessly from ICIE FOV and connect to the heliospheric imager FOV |
| **Heliospheric Imager with Polarization (HIP)** *STEREO, PUNCH* New: Polarization capability | To track solar features into the heliosphere, their impact on Earth, provide line-of-sight electron column densities for FETCH analysis |
| **Faraday Effect Tracker of Coronal and Heliospheric structures (FETCH)** New instrument | To determine the magnetic field structure and evolution of solar wind structures in the Sun-Earth connected space |
| **Radio and Plasma Wave instrument for MOST (M/WAVES)** *STEREO* New: Improved antennas to minimize dust impact | To track shocks and electron beams from Sun to 1 au, determine the source region configuration of type III storms and the implications of seeds particles accelerated at the storm source |
| **Solar Wind Plasma Instrument (SWPI)** *Rosetta, SWFO-L1* New: CME speeds up to 2500 km/s | To infer solar magnetic structures at 1 au and CIR evolution |
| **Solar Wind Magnetometer (MAG)** *Parker Solar Probe* | To infer solar magnetic structures at 1 au, CIR evolution |
| **Solar High-energy Ion Velocity Analyzer (SHIVA)** *CeREs, CUSP CubeSats, Van Allen Probes* New: Proton energy channels up to 500 MeV | To determine spectra of electrons, and ions from H to Fe at multiple spatial locations and use energetic particles as tracers of magnetic connectivity |

[a]Missions listed in italics provide the heritage for the MOST payload

## 3.2 The Magnetic and Doppler Imager (MaDI)

The Magnetic and Doppler Imager (MaDI) will measure the photospheric/chromospheric magnetic and velocity fields, map the photospheric magnetic field and help study the magnetic field (active region) evolution and its connections to physical conditions in the tachocline through seismology (see e.g., Christensen-Dalsgaard, 2021 and references therein). The Doppler images from MaDI at L4 and L5 can be combined with those obtained by similar instruments on the Sun-Earth line (ground-based and L1) for probing the whole convection zone (e.g., Bemporad 2021). Active regions can be tracked from the frontside to backside with improved accuracy (e.g., Yang et al. 2023). The advantage of multiview is that two of the three field components of both velocity and magnetic fields, can be obtained over common area just using the line-of-sight component. Further, we can resolve the ambiguity in field directions in the overlapping area between two magnetographs. Magnetograms obtained from L5 view can help space weather forecasting by observing active regions and coronal holes well before they rotate to the Earth view (Gopalswamy et al. 2011a; Kraft et al. 2017). Magnetograms from L4 view will provide a more direct view of SEP source regions (Posner et al. 2021). A significant fraction of SEP events, which affect near-Earth environment originate from active regions near or even behind solar west limb as observed from Earth. Having SEP source region observations from L4 would also allow extending a so-called safe zone for spacecraft travelling to Mars. Here, the safe zone is an area in heliosphere covered by a robust modeling of space weather. Surface magnetic field measurements from all vantage points will extend the coverage of more than ~70% of the entire solar surface compared to less than a quarter of the surface at present (Pevtsov et al. 2020; Bemporad 2021). Combined observations from L5, L1/Earth, and L5 will also significantly improve the visibility of solar poles. The visibility is improved since the poles are hidden at different times of the year from different views from the ecliptic. For example, in Earth/L1 view, the south pole is hidden for six months centered on September. From L5 and L4 views, the south pole is hidden for six months centered on July and November, respectively. The net result is that each pole is hidden for only one month in a year (in September for south pole and March for north pole). Thus, the solar poles will be observed most of the time (albeit at small angles) by combining MOST and L1/Earth views, providing significant improvement in modeling the solar wind and CMEs (Posner et al. 2021).

Magnetographs such as the Michelson Doppler Imager (MDI, Scherrer et al. 1995) on SOHO or the Helioseismic and Magnetic Imager (HMI; Scherrer et al. 2012) on the Solar Dynamics Observatory (SDO) are traditional instruments with complex optical systems. There have been efforts to reduce the SWaP (Size Weight and Power) of these traditional designs including the Photospheric Magnetic Field Imager (PMI, Staub et al., 2020) based on Solar Orbiter's Polarimetric and Helioseismic Imager (SO/PHI, Solanki et al. 2020) design, and the Compact Magnetic Imager (see Fig. 2 from Hurlburt and Berger, 2021) based on HMI design.

More recently, Compact Doppler Magnetograph (CDM) instrument (Hassler et al., 2022; Gosain et al., 2022) based on Global Oscillations Network Group instrument (GONG, Harvey et al., 1996; Hill, 2018) was proposed for the Solaris mission (Hassler et al., 2020). CDM is demonstrated to be TRL6 with a mass estimate of only 16 kilograms with 20% margin (Hassler et al., 2022). CDM uses innovative design where a group of three solar lines is used to increase the signal-to-noise ratio (SNR) of the measurements while simultaneously providing immunity to measurements from large spectral shifts resulting from high spacecraft velocity.

While the aforementioned traditional instrument designs are well understood and have proven very successful in Doppler-magnetography, the mass constraints typical of deep space missions requires exploring new alternative designs with total mass of only few kilograms. One such technology for MaDI is based on the photonics chips and is described below.

MaDI will make use of the latest developments in magnetography based on recent progress in photonics and electronics. The Imaging Photonic Spectropolarimeter for Observing the Sun (IPSOS, Hurlburt, Vasudevan and Chintzoglou, 2022) shown in Figure 2 is based on the recently demonstrated laboratory prototype (Hurlburt, 2021; Hurlburt et al., 2023) where the bulk of optical elements are contained in a single multilayer wafer instead of the traditional mechanical filter components. Instead of using a telescope to guide the solar image through a spectropolarimeter (SP), IPSOS first feeds the solar signals into an array of heterodyne SPs on a photonic integrated chip (PIC) fed by a tunable laser. The laser also maintains coherence between the relative phases of the SPs. The outputs of the SPs are then combined computationally to create a magnetogram. The optical package is reduced to a single wafer while the electronics exploit compact, low-power RF Systems on a Chip (RFSoCs). Given the deep-space locations of the MOST instruments, it is advantageous to have reduction in cost, mass, and risk. The IPSOS concept provides these reductions because there will be no assembling of major components; instead, the components will be printed using standard lithographic techniques.

The IPSOS instrument meets all MaDI's requirements with a spatial resolution and cadence similar to SOHO/MDI. IPSOS has a maximum baseline of 18 cm which fits on a standard 8-inch silicon wafer. The speed at which we can collect sufficient u-v samples drives the number of apertures and the overall power and mass of the instrument. Table 3 displays the projected SWaP of the IPSOS instrument for MOST while Figure 2(a) shows what it would look like if built today (IPSOS1) with 20 kg of mass and 47 liters of volume. The second-generation build will have a mass < 2 kg with a volume of only 2 liters (IPSOS2). Since the SWaP of the optical components are negligible, IPSOS can easily support multiple arrays for different spectral bands. The two additional apertures are tuned to capture data in the 1083 nm He I and 854 nm Ca II chromospheric lines.

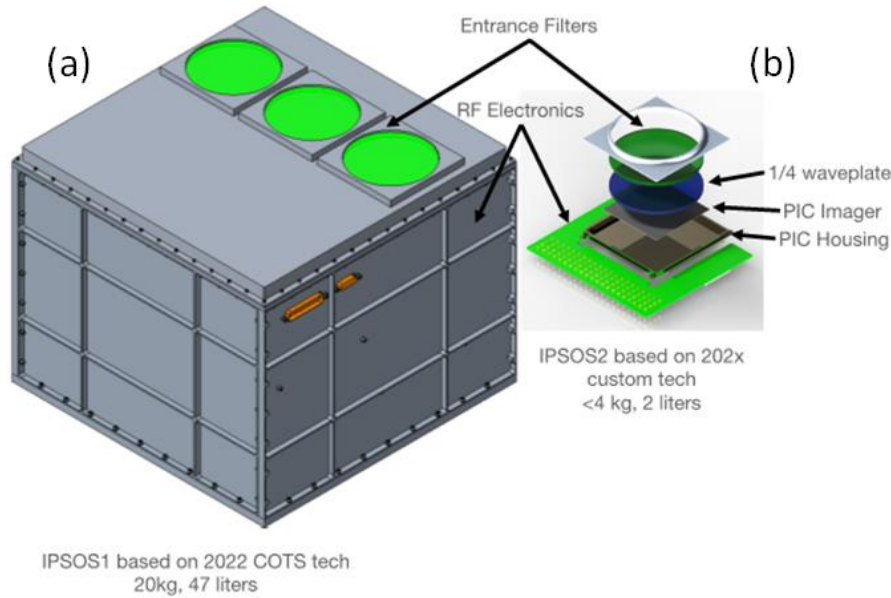

**Figure 2**. (a) First- and (b) second-generation concepts for the Imaging Photonic Spectropolarimeter for Observing the Sun (IPSOS). This version uses 9 cm apertures which can be scaled up to 18 cm to meet MaDI requirements.

Following generations will be even lower SWaP as technology matures, leading to the wafer-like vision in Figure 2b. Data products will include magnetic and Doppler imaging in the photosphere and chromosphere.

**Table 3**. IPSOS Instrument Characteristics

| Parameter | Value | Comment |
|---|---|---|
| Mass | 6 kg | Estimated |
| Volume | 6 liters | Estimated |
| Average Power | 20 W | Estimated |
| Real-time Data Rate | 0.14 Mbits/sec | Requirement |
| Field of View | 53 arc-min | Requirement |
| Maximum baseline | 18 cm | Required for spatial resolution |
| Measurement Type | Solar Magnetic Fields and Doppler Velocity | Spatial scale of 1.0 arc-sec/pixel. SNR > 2000. |
| Measurement Wavelength | 1564.8 nm | Fe I 1564.8 nm, He I 1083 nm, and Ca II 854.2 nm |
| TRL | 4 | Based on MICRO HTIDeS project |

### 3.3 The Inner Coronal Imager in EUV (ICIE)

Inner Coronal Imagers such as Hinode's X-ray Telescope (XRT, Golub et al., 2007), SOHO's Extreme-ultraviolet Imaging Telescope (EIT, Delaboudiniere et al., 1995) and SDO's Atmospheric Imaging Assembly (AIA, Lemen et al., 2012) have demonstrated the importance of observing the Sun close the solar surface. SOHO's CME watch program that combined EIT images with the Large Angle and Spectrometric Coronagraph (LASCO, Brueckner et al., 1995) images have contributed enormously to understanding the early evolution of CMEs (Dere et al., 1997; Gopalswamy and Thompson, 2000). STEREO's Extereme Ultraviolet Imager (EUVI, Wuelser et al. 2007) further confirmed the usefulness of EUV images in understanding the 3-D structure of quiescent transient coronal structures and their relation to heliospheric structures such as CMEs and SIRs. Recent instruments such as the Sun Watcher using Active Pixel system detector and image processing (SWAP, Berghmanns et al., 2006), the Solar Ultraviolet Imager (SUVI, Seaton and Darnel, 2018), and the Full Sun Imager (FSI) telescope of the Extreme Ultraviolet Imager (EUI) on board Solar Orbiter (Berghmanns et al. 2023) have demonstrated that the extended corona can be imaged in EUV in a much wider field of view (FOV, out to ~3 Rs). ICIE extends the wide FOV EUV imager and the design is similar to ISS Coronal Spectrographic Imager in the EUV (COSIE, Golub and Savage, 2016). ICIE will identify: coronal structures from solar limb/disk into the coronagraph FOV, changes in open field connectivity from 1-3 Rs, streamer plasma inhomogeneities, filaments, CMEs, EUV waves/shocks, coronal dimmings, and current sheets associated with CMEs.

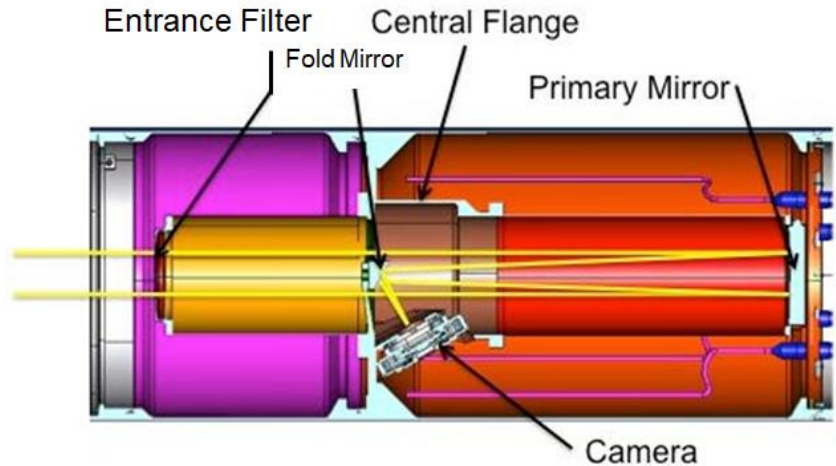

**Figure 3**. Optical design of the ICEI instrument showing the entrance filter, primary mirror, fold mirror and the camera.

Figure 3 shows the ICIE optical design. It is a compact (70 cm × 20 cm × 20 cm) light-weight (~40 kg) design with a passband in the wavelength range 17.1 to 20.5 nm. ICIE will use a CMOS camera (3k × 3k, 10 μ pixels).

### 3.4 The Hard X-ray Imager (HXI)

The hard X-ray Imager (HXI) investigates solar flares by providing diagnostics of the hottest (>8 MK) flare plasmas and flare-accelerated electrons above 10 keV. The hard X-ray images help clarify the flare structure thought to be associated with the post-eruption arcade (PEA) in EUV. The two views provide more opportunities to observe loop-top hard X-ray sources (Masuda et al., 1994). The HXI design is based on the Spectrometer/Telescope for Imaging X-rays (STIX) on board Solar Orbiter (Krucker et al., 2020, and references therein). There is no major difference between HXI and STIX but the use of two views from L4 and L5 will help obtain the 3-D structure of flare structures and their relation to core dimming and PEA observed in EUV.

Figure 4 shows the STIX design to be adapted for HXI on MOST. HXI consists of three major elements from the front to the back of the instrument: (i) a pair of X-ray transparent entrance windows, (ii) the imager consisting of two widely separated grids for Fourier-transform bigrid imaging, and (iii) Detector Electronics Module containing electronics and cadmium telluride detectors, and an X-ray attenuator. Details on the instrument can be found in Krucker et al. (2020).

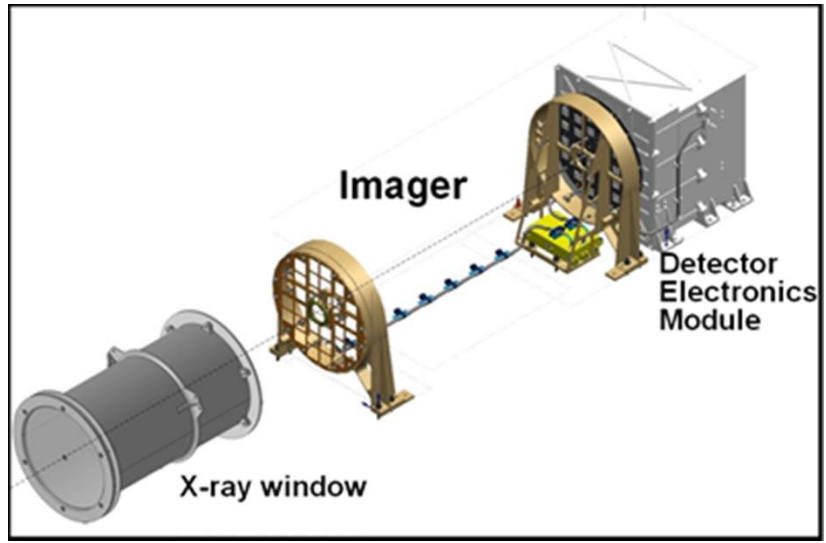

**Figure 4**. The HXI design based on Solar Orbiter's STIX showing the three major components: X-ray window, imager, and the detector elecronic module. (Adapted from Krucker et al., 2020).

### 3.5 The White-light Coronagraph (WCOR)

The White-light Coronagraph (WCOR) will build upon the success of SOHO and STEREO coronagrahs by improving the instrument with recent technology. White-light coronagraphs have become a key instrument in heliophysics investigations because of their ability to image the extended solar atmosphere using Thomson-scattered photospheric light (e.g. Koutchmy, 1988). Coronagraphs provide the essential observations of the structure and dynamics of the outer corona and near-Sun interplanetary medium. WCOR will obtain polarized and total brightness images of the Sun's corona with a FOV in the heliocentric range of 2 to 15 Rs. WCOR data determine 3-D geometry, morphology, kinematics and mass of expanding CMEs and provide global configuration of the outer corona.

The design of WCOR is based on the science objectives and key measurement requirements discussed above. The optical and mechanical designs are shown in Figure 5. WCOR is designed to provide a similar but improved performance from Sun Earth Connection Coronal and Heliospheric Investigation (SECCHI) COR2 (Howard et al. 2008) (and the Baloon-borne investigation of the temperature and speed of the electrons in the corona (BITSE, Gopalswamy et al., 2021). The improvements include: (1) reduced external occulter (EO) cutoff, which reduces the vignetting for the field near the inner edge of FOV; (2) an internal occulter is added to further improve the diffraction suppression in comparison with BITSE; (3) a larger format polarization detector array is used to not only eliminate the need for a polarization wheel mechanism, but also to capture all polarization information simultaneously. The polarization detector overcomes the image smear introduced by wheel-based polarization mechanism and has been sucessfully used in BITSE.

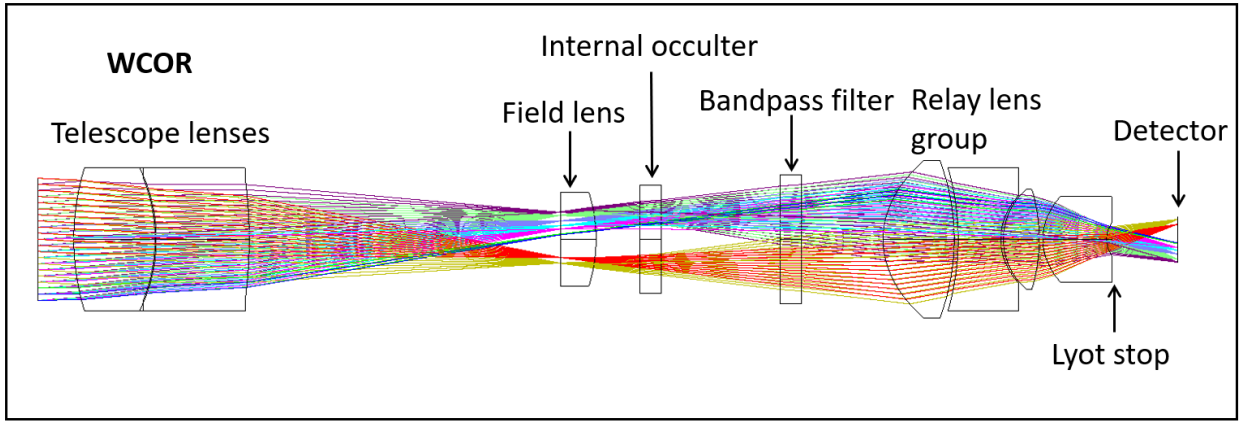

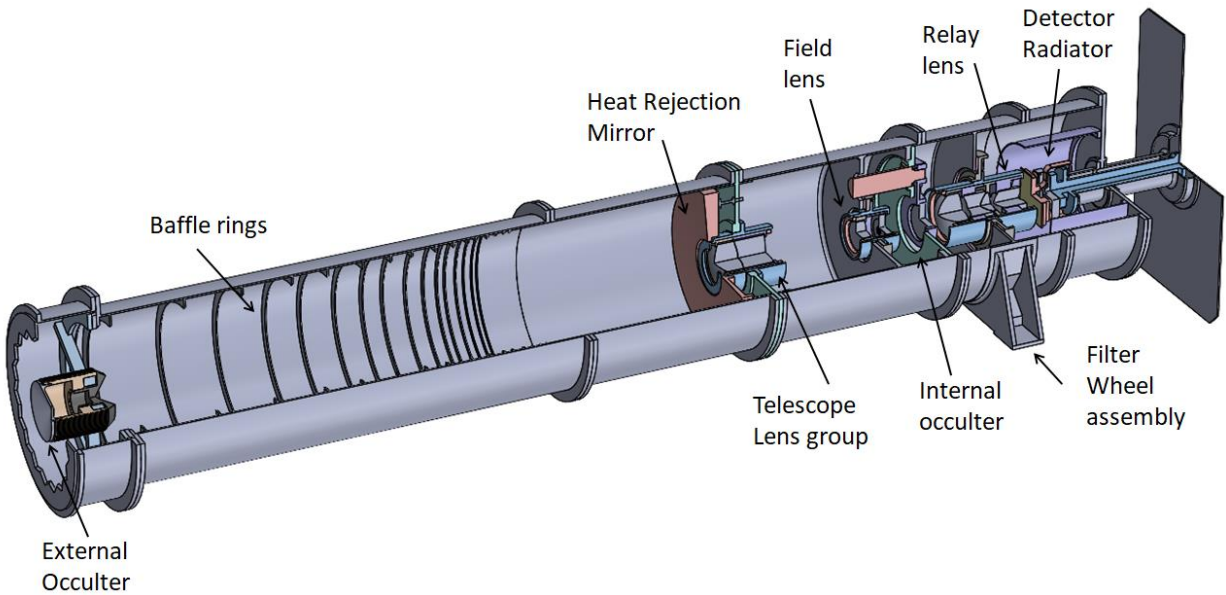

**Figure 5**. The optical (top) and mechanical (bottom) designs of WCOR. The coronagraph is externally occulted, the occulter being a threaded, tappered right frustum. The filter wheel is to switch between broadband (650-750 nm) and H-alpha (656 nm) filters. The overall length of the coronagraph is ~1 m. The detector is a 4k × 4k CCD with 10 μ pixels.

**Table 4.** WCOR specifications

| Parameter | Value |
|---|---|
| Pixel size (μm) | 10 |
| Detector (Teledyne E2V) | 4k x 4k |
| Chip size (mm) | 40 x 40 |
| FOV (Rs) | 2.5-15.0 |
| FOV (°) | 4 x 4 |
| Effective focal length (mm) | 273.7 |
| Plate scale ("/Super pixel, 20 µm) | 15 |
| Wavelength range (nm) | 650 - 750 |
| EO inner cutoff (Rs) | 2.0 |
| IO inner cutoff (Rs) | 2.5 |
| Distance (A0 - A1) (mm) | 600 |
| A1 diameter (mm) | 34 |

WCOR specifications are given in Table 4. Use of the polarization detector results in a spatial resolution of 15"/super pixel (2 x 2 polarizer bin). The Nyquist resolution is 30", as in COR2. The modulation transfer function is above 0.5 except for the region of the FOV severely vignetted by the EO. The diffraction brightness B (relative to the mean solar brightness Bs) at 2.5 Rs is ~4 × $10^{-9}$, which is about an order of magnitude lower than the F-corona brightness. The SNR analysis considered the external brightness from K-corona, F-corona, diffraction, and internal scattering, and internal brightness from read noise and dark current responsible for generating the total photoelectrons per second in the CCD to fill 80% of the full-well depth of 10,000 electrons. It was determined that at 3 Rs the integration time that satisfied the above requirement was 6 seconds with a SNR of 23 in the brightest super pixel (aligned with tangential K-corona) and 5 in the least bright super pixel (aligned with radial K-corona).

### 3.6 The Heliospheric Imager with Polarization (HIP)

Heliospheric imaging pioneered by STEREO/SECCHI has revolutionized our understanding of the large-scale structure of the inner heliosphere (Socker et al., 2000; Howard et al. 2008; Harrison et al. 2009; Amerstorfer et al. 2021). The Heliospheric Imager (HI) instrument has been successfully used in several missions such as STEREO (Eyles et al., 2009), Parker Solar Probe's Wide-Field Imager for Solar Probe (WISPR, Vourlidas et al., 2016), and the Solar Orbiter Heliospheric Imager (SoloHI, Howard et al., 2020). The Wide Field Imager (WFI) currently under development to be flown on the Polarimeter to Unify the Corona and Heliosphere with Polarization (PUNCH, DeForest et al., 2022) has added a new dimension to heliospheric imagers: polarization. The Heliospheric Imager with Polarization (HIP) will follow the design reported in Lavraud et al. (2016). In addition to the polarization capability, HIP will have better sensitivity and a steady view from L4 and L5. The better sensitivity will help distinguish between the flux rope and shock in fast CME events at large distances from the Sun. Polarimetry is critical to identifying feature chirality and substructure, and to tracking event trajectories in 3-D; and adds precision to overall background subtraction, improving line-of-sight (LOS) density estimates, which are important for Faraday-rotation measurements of the interplanetary magnetic field (IMF) using HIP and FETCH together (DeForest et al. 2022).

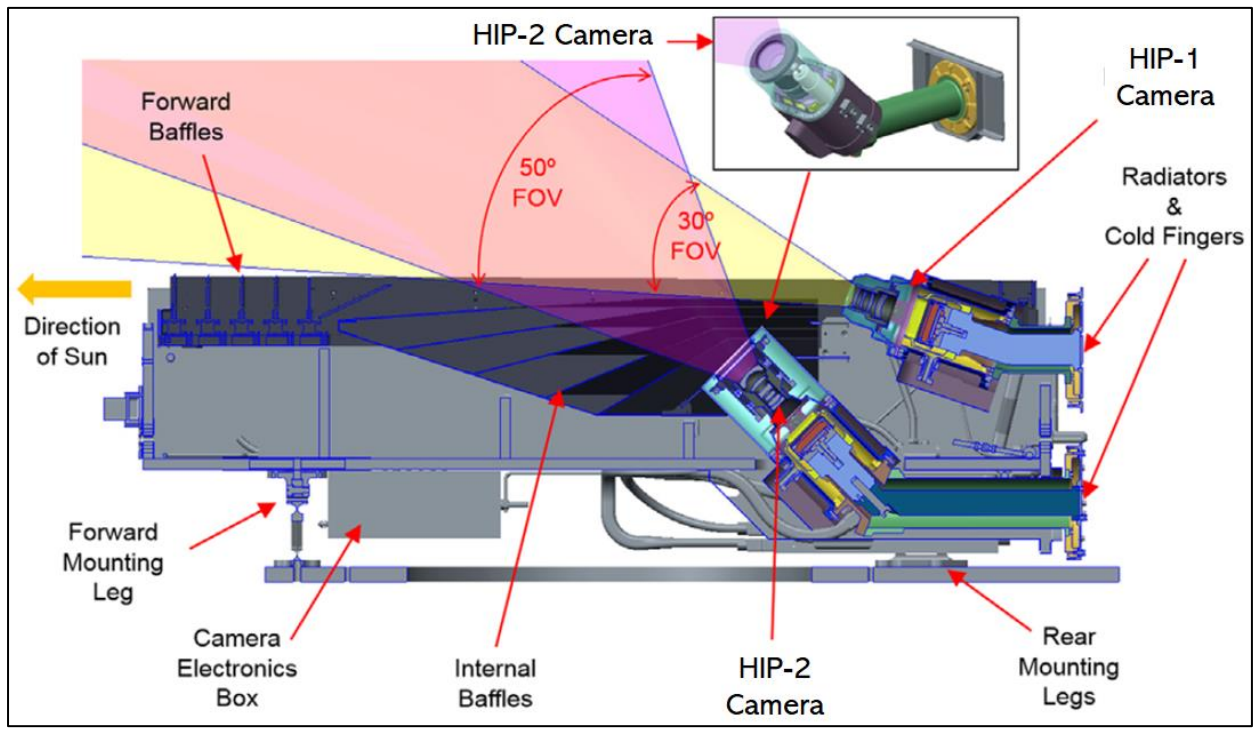

**Figure 6**. A schematic showing the two cameras HIP-1 and HIP-2 with 30º and 50º FOV, respectively. Key subsystems such as baffle systems, electronic box, and radiators are noted. The cameras will use 2k x 2k detectors similar to those being developed for PUNCH. The spatial resolution is ~4 arcmin and the image cadence is ~10 min. [Adapted from Lavraud et al., 2016].

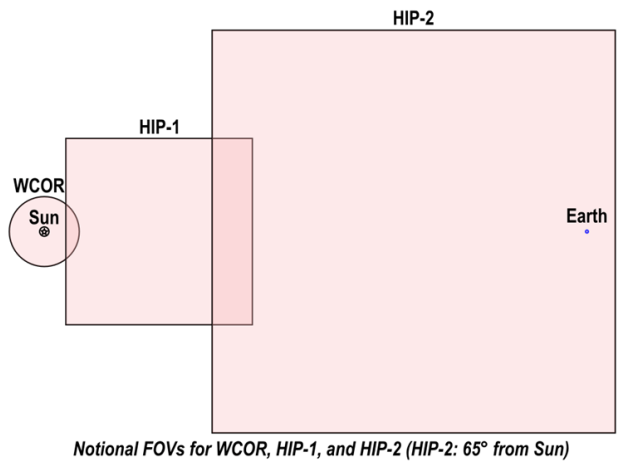

*Notional FOVs for WCOR, HIP-1, and HIP-2 (HIP-2: 65° from Sun)*

**Figure 7**. Overlapping FOVs of HIP-1 and HIP-2 with the WCOR FOV. HIP will do polarized heliospheric imaging from 10-20 Rs to 65°from the Sun. The HIP-2 FOV extends beyond Earth.

HIP is environed to have a wider FOV to include Earth within the HOP FOV. Figure 6 shows the two cameras (HIP-1 and HIP-2) similar to the STEREO counterparts with a combined FOV whose outer boundary is beyond Earth and inner boundary overlapping with the coronagraph FOV (see Figure 7). HIP makes use of extensive heritage from the polarimetry, background subtraction, and post-processing techniques developed for PUNCH, and the HIP data pipeline provides polarimetric background-subtracted images of the solar wind.

## 3.7 The Faraday Effect Tracker of Coronal and Heliospheric structures (FETCH)

Faraday effect refers to the rotation of the the plane of polarization of a linearly polarized wave traveling through a magnetized plasma (Collett, 1992). The extent of rotation ($\Delta\chi$) depends on the electron density distribution ($n$) and magnetic field ($\boldsymbol{B}$) component along the line of sight ($s$) in cgs units:

$$\Delta\chi = \lambda^2 \left(\frac{e^3}{2\pi m_e c^4}\right) \int_0^S n(s)\boldsymbol{B}(s).d\boldsymbol{s} = \lambda^2 RM, \qquad (1)$$

where $\lambda$ is the wavelength employed RM is the rotation measure determined by the line-of-sight integration of the product $nB$. Thus, with measuring the FR angle, we can invert equation (1) to estimate the density and magnetic field along the line of sight. Since the HIP field of view overlaps with the spatial domain where FETCH makes measurements, one can obtain the line-of-sight intergated density from HIP observations independent of magnetic field, so that magnetic field structure can be deduced. The technique is well known and has been extensively used in the past to observe FR of signals from distant radio sources and from spacecraft (see Kooi et al., 2022, and references therein).

FR measurements at 1 au have shown to be an excellent tool for inferring the magnetic field of the solar corona, including CMEs (Mancuso and Garzelli, 2013) and the background solar wind (Bird, 2007). Detailed knowledge of the magnetic field content of solar transients such as CMEs and CIRs as they propagate along the Sun-Earth line, is crucial for effectively forecasting space weather. The estimated CME magnetic field and its orientation, well before it reaches 1 AU, can be used to determine the geo-effectiveness of the CME. Many background radio sources emitting linearly polarized signals can help determine the FR in CMEs (Howard et al., 2016). However, the existing methods for measuring FR through a CME is currently limited to ground-based measurements with powerful radio telescopes such as Very Large Array (VLA). Even though FR measurements have been made for decades for probing the solar wind (and CMEs) to obtain the plasma densities and magnetic fields using external linearly polarized sources in the sky, the multiple LOSs observations envisioned to be performed by MOST are yet to be carried out. Liu et al. (2007) demonstrated a method to measure the magnetic field orientation of CMEs using FR measurements. The authors proposed time-dependent FR mapping for calculating CME propagation away from the Sun to resolve its geometry. Jensen and Russell (2008) showed that by fitting a force-free flux rope model to observations from a spacecraft, one could obtain various information about the flux rope, such as its orientation, size, and velocity. Combining this with electron density measurements (see section 5), one can obtain the magnetic field strength. These authors also emphasized on the need of multiple LOSs FR observations to obtain the proper flux rope geometry and remove structural ambiguity.

Bird (2007) utilized satellite signals as background radio sources, however, these observations were also carried out with large ground-based radio telescopes. Ground-based radio observations are affected by ionospheric plasma, which introduces additional FR on its own. Furthermore, because the Sun is a bright source of radio emission, only a few powerful antennas such as the Green Bank Telescope or VLA are capable of viewing distance radio sources on the flank of the expanding CME structure. The structures behind the nose of the CME cannot be sampled by directly pointing at the Sun. FETCH will overcome these shortcomings by transmitting and receiving in space.

While Faraday rotation provides information about column-integrated density and magnetic field, recent state-of-the-art modeling of FETCH observations demonstrates that information about the distribution of these parameters could be derived (see, e.g., Jensen et al. 2023; Wexler et al. 2023). The flow of the plasma across the FETCH LOS consists of a small offset in sampled plasma between simultaneously counter crossing signals between two spacecraft. They take as long as 16 minutes to traverse between MOST 3 & 4 for example, sufficient time for a half-solar radii difference to develop; the scale size of the radio observation column is a tenth of this. This difference in the total electron content is measurable and can be used in a coarse tomographic analysis of the resulting time series. Tomography can also be used between the four lines-of-sight via constraining an MHD model or with simplifying assumptions regarding plasma outflow and structural coherence characteristics.

Spacecraft-to-spacecraft FR was first demonstrated by using the radio transmissions from the Radio Plasma IMAGER (RPI; Reinisch et al., 2000) on the IMAGE satellite to the Wind and Cluster satellites (Cummer et al., 2001; 2003). The distance scales in the experiment were up to 15 Earth radii in the Earth's magnetosphere, which is several orders of magnitude smaller than the >1 au distances over which FETCH will make measurements.

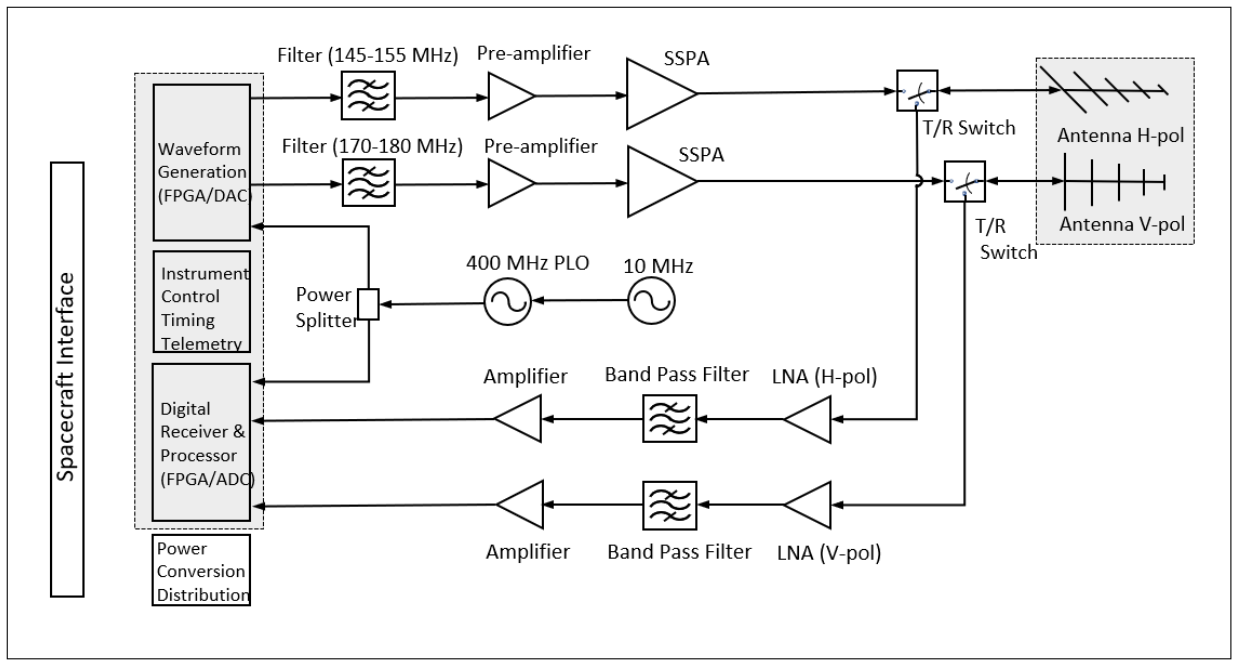

**Figure 8**. FETCH block diagram showing notional digital backend electronics design and signal transmission-reception scheme. FETCH will transmit in a single polarization (vertical V or horizontal H) and receive at both polarizations simultaneously while the transmitter is off. Transmission will be in one polarization at 165 MHz and the other at 225 MHz. Receiving will be at both polarizations and frequencies.

The main FETCH subsystems for the MOST mission are shown in Figure 8. The current baseline FETCH design includes a log-periodic dipole antennas operating at two frequencies: 165 and 225 MHz. The antenna will have a length of 3.3 m and a maximum width of 1 m. The antenna can be stowed into a canister for launch and then deployed. The transmitted FETCH signal will be chirp compressed to boost the signal gain (Bernfeld et al., 1965). A detailed list of parameters for the FETCH system is given in Table 5 including the transmitter (Tx) and receiver (Rx) elements.

**Table 5**. List of FETCH system parameters

| FETCH System Parameters | | |
|---|---|---|
| Parameters | Values | Note |
| Freuqency (MHz) | 165.0, 225.0 | |
| Tx Peak Power (W) | 200.0 | |
| Tx Chirp Pulsewidth (s) | 2.0 | |
| Tx Bandwidth (Hz) | 1000.0 | |
| Dutycycle (%) | up to 50.0 | |
| Antenna Base Element Size (m) | 1.0 | Largest dipole element dimension of LPDA |
| Antenna Boom Length (m) | 3.3 | Dimension along boresight |
| Antenna Beamwidth(deg) | ± 30.0 | |
| Antenna Gain (dB) | 10.5 min | |
| Integrated Cross-pol Isolation (dB) | 40.0 | |
| Rx Bandwidth (Hz) | 1000.0 | |
| Rx Noise Figure (dB) | 2.0 | |
| Integration Time (s) | 100.0 | |
| Signal to Noise Ratio (dB) | 8.0 | |
| | | |
| Data Rate (kbps) | 100.0 max | |
| Power (W) | 515.0 | Assuming two 200 W SSPAs with 50% efficiency, 50% Tx dutycycle and 10% contingency |
| Weight (kg) | 88.0 | Assuming 10 kg for deployable antenna and 20 kg for antenna boom |

### 3.8 The Radio and Plasma Wave instrument for MOST (M/WAVES)

Solar radio emission at frequencies below the ionospheric cutoff cannot be observed from the ground. These frequencies contain most important information on eruptive phenomena and the interplanetary medium through which solar disturbances propagate (Cane et al. 1987; Gopalswamy 2011). The radio and plasma wave experiment (WAVES, Bougeret et al., 1995) onboard the Wind spacecraft demonstrated the importance of observing at all frequencies below the ionospheric cutoff down to ~20 kHz (i.e., from decameter to hectometer to kilometer wavelengths). The lowest frequency corresponds to the local plasma frequency at the observing spacecraft, while the highest frequency corresponds to ~2 Rs (the middle corona). The advent of the frequency range 1-14 MHz by Wind/WAVES resulted in a number of discoveries, especially because of the coronagraph images provided by SOHO in the overlapping spatial domain (see Gopalswamy, 2011, for a review). STEREO/WAVES (Bougeret et al., 2008) provided similar spectral coverage with a slightly higher upper cutoff (~16 MHz) and different antenna system (three monopole stacer antennas, Bale et al., 2008) on each of the two STEREO spacecraft. The major advantage of the two views is that triangulation can be used to identify the location of a shock or electron beam emitting radio waves (Krupar et al., 2012; Mäkelä et al., 2016). WAVES observations help track type II radio bursts, type III radio bursts, and radio noise storms that provide information on the disturbances as well as the magnetic and density structure of the heliosphere.

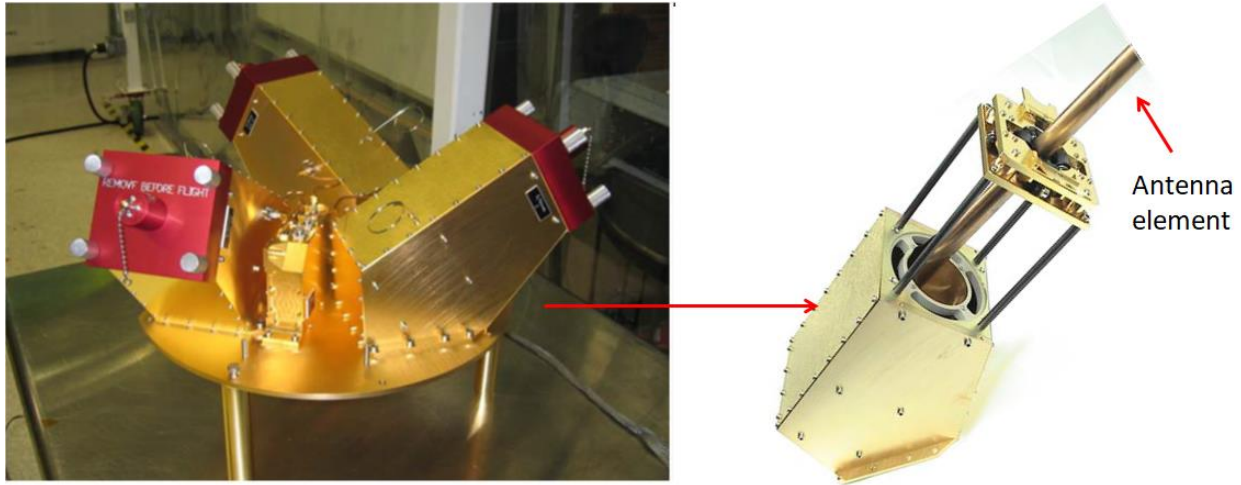

**Figure 9**. (left) Stowed configuration of the three mutually orthogonal antenna elements and pre-amplifier enclosure. The red block is the pre-deployment retaining cover. (right) one of the antenna elements in the deployed state (only part of the 6-m antenna element is shown) [Adapted from Bale et al., 2008; Bougeret et al., 2008]

The MOST/WAVES (M/WAVES) experiment will closely follow the design of S/WAVES. Figure 9 shows the three mutually orthogonal antenna elements to be used for M/WAVES: in the stowed configuration with one of the antennas deployed. The antennas will make electric field radio measurements from 10 kHz to ~25 MHz in at least 3 channels. Plasma waves measurements of the quasi-thermal noise spectrum with < μV sensitivity and $\Delta f/f < 4\%$ spectral resolution to resolve the electron plasma frequency and thermal plateau. Rapid waveform measurements of individual antenna voltages will be made. These will be useful in characterizing plasma waves associated with SEP electron events and dust impact signatures on the MOST spacecraft.

The only improvement in M/WAVES is that the antenna elements will be redesigned to reduce the surface area to minimize the effect of dust impact. Another improvement is the increase in sensitivity at low frequencies, so the plasma line due to the quasithermal noise can be observed better to obtain plasma density in the vicinity of the observing spacecraft.

### 3.9 Solar Wind Plasma Instrument (SWPI)

The Solar Wind Plasma Instrument (SWPI) is based on the Ion and Electron Sensor (IES, Burch et al. 2006) that completed operations successfully on the Rosetta mission. SWPI has a compact dual measurement sensor to measure ion and electron velocity distribution functions. Figure 10 shows the main components of IES along with the instrument block diagram. Particles enter the grounded entrance grid and are deflected according to energy- and elevation angle-dependent curved bipolar deflector electrodes into field-free apertures. Particles then enter the top-hat region Electrostatic Analyzer (ESA) segments and get focused onto microchannel plates (MCPs), with delay line anodes.

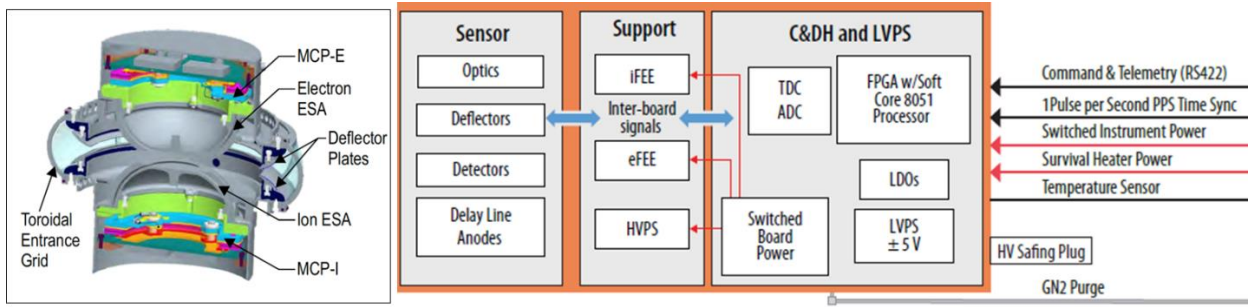

**Figure 10**. Cross-sectional view of the SWPI sensor (left) and the instrument block diagram (right). The sensor-head is made up of the common entrance grid and deflector plates, as well as electro-static analyzers (ESA), micro channel plate (MCP) detectors and front-end electronics (FEE) for electrons and ions. The sensor head is attached to an electronics box (Ebox). The Ebox uses a low voltage power supply (LVPS) to distribute incoming power from the spacecraft to all instrument components. A command and data handling board that includes an FPGA with embedded flight software controls the subsystems and processes data from the detectors. The high voltage power supply (HVPS) provides high voltage to the ESAs, deflectors and MCPs. It also includes an FPGA that uses tables to control sweeping of the ESA and deflector voltages necessary to detect particles across the angular and energy ranges.

The Sensor heritage is from Rosetta IES, while the Electronics are based on the Solar Wind Plasma Sensor (SWiPS) instrument on the NASA/NOAA SWFO-L1 mission and Magnetic Anomaly Plasma Spectrometer (MAPS) on the Lunar Vertex lander mission. Flight model builds for both instruments are underway with completion expected in summer 2023.

**3.10 Solar Wind Magnetometer (MAG)**

For the accomplishing the MOST scientific objectives, it is critical to measure the magnetic fields of CMEs, shocks, and other magnetic field structures reaching 1 au. The magnetic field indicates the structure of the CMEs arriving at the spacecraft at L4 and L5, including the differences between the two spacecraft locations. The magnetic field also indicates the magnitude variation of the shocks arriving at the spacecraft, including the difference between the two locations. The three-coordinate vector measurements of the magnetic field indicate changes that have occurred relative to the Parker Spiral. Measurements of ion and electron density and solar wind speed provide additional information about the magnetic field structures. CIRs arrive at L5 first, then at Earth, and finally at L4, thus helping understand the evolution of CIRs.

The MAGs on MOST will be duplicates of the Parker Solar Probe (PSP) magnetometers, which are part of the FIELDS experiment (Bale et al., 2016). PSP MAGs are triaxial fluxgate magnetometers built by Goddard Space Flight Center and similar to those successfully flown aboard MAVEN, Van Allen Probes and GOES-18. The PSP MAGs operate at a maximum cadence of 297.97 samples/sec. Four different dynamic ranges provide a full-scale resolution of +/- 1024 nT, +/- 4096 nT, +/- 16384 nT and 65536 nT, determined by the ambient magnetic field. The smaller dynamic ranges provide smaller sampling resolution, starting with 0.03125 nT/ADU in the +/- 1024 nT range and 0.125 nT/ADU, 0.5 nT/ADU and 2.0 nT/ADU in the respectively larger dynamic ranges. PSP MAGs are functioning quite well providing numerous observations of Alfvenic magnetic field switchbacks (i.e., Kasper et al., 2019; Bale et al., 2021), and observations of the structure of the near-Sun space magnetic field (Bale et al., 2019). The MAG instruments for MOST will be near identical and build-to-print, with only minimal changes due to spacecraft accommodations, making them very cost effective and low risk. Figure 11 shows the PSP MAG sensor. MOST will use two magnetometers, one inboard and one outboard. This would give us one more calibration technique using the two magnetometers to disentangle the contribution of the spacecraft and is similar to how PSP, MAVEN, Juno and GOES have flown recently.

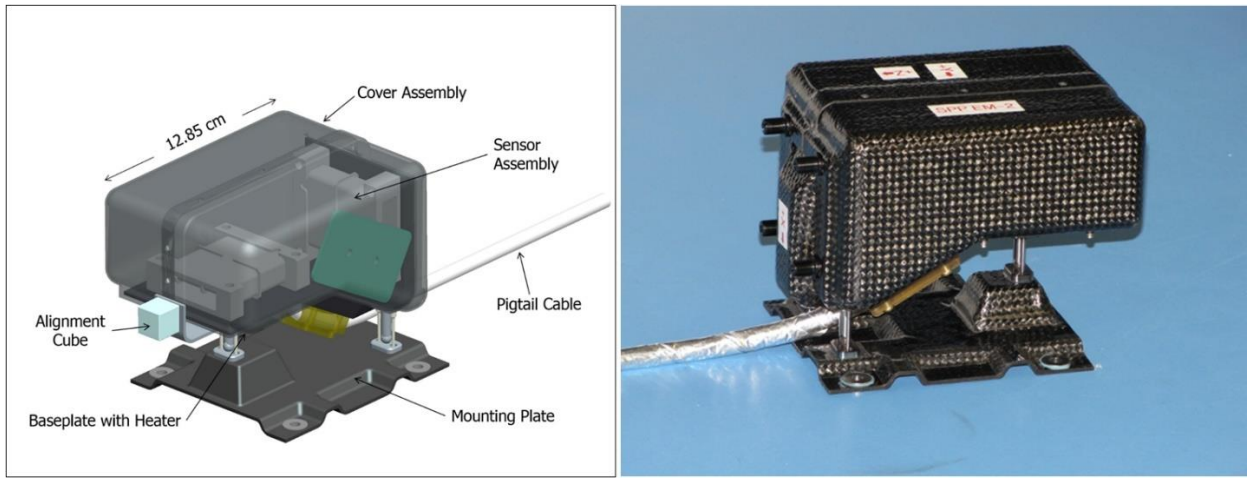

**Figure 11**. (left) Drawing of the two PSP MAG sensors, showing the sensor assembly. (right) Actual photograph of one of the PSP MAGs. As in PSP MAGs, MOST MAG will have a data rate of ~256 samples per second.

### 3.11 Solar High-energy Ion Velocity Analyzer (SHIVA)

Solar High-energy Ion Velocity Analyzer (SHIVA) is MOST's energetic particle detector needed to understand the origins, energization, and transport of charged particles from the Sun and inner heliosphere. SHIVA will characterize: the energy spectra, event classes, longitudinal features, and composition (from He to Fe) of SEP events. Furthermore, SHIVA will also characterize SEP electrons from ~tens of keV to ultra-relativistic energies and Anomalous Cosmic Rays (ACRs); the latter will advance our understanding of solar influences on the distant heliosphere. Additional science addressed by SHIVA include measurement of the variability of primary galactic cosmic rays and Forbush decreases. Finally, SHIVA will help monitor the radiation environment **in** the inner heliosphere, an important space weather contribution.

Figure 12 shows a schematic of the SHIVA instrument. SHIVA comprises two sensor stacks: a detector stack made of solid-state detectors (SSDs) behind space-facing avalanche photodiodes (APDs). The APD-SSD combination enables measurement of electrons from ~20 keV to ~5 MeV; protons from ~200 keV to ~100 MeV; heavier ions (He to Fe) from 2 to 200 MeV/nuc in multiple differential energy channels. The energy range can be extended using individual SSD pulse height analysis (PHA), e.g., up to 500 MeV for protons. The energy resolution is <30% and the time resolution is software selectable, typically ~1 min. SHIVA will closely follow the design of the Miniaturized Electron pRoton Telescope (MERiT), which is a low-mass, low-power, compact instrument using an innovative combination of particle detectors, sensor electronics, and onboard processing. MERiT flew on the Compact Radiation belt Explorer (CeREs), a 3U CubeSat in the ~500 km low Earth orbit (LEO, Kanekal et al., 2019) and the CUSP (CubeSat to measure Solar energetic Particles) CubeSat.

Another version of MERiT is under development for the Heliophysics Environmental and Radiation Measurement Experiment Suite (HERMES) on the Lunar Gateway. In addition to CeREs, CUSP and HERMES, SHIVA has heritage from the Relativistic Electron Proton Telescope (REPT) instrument on Van Allen Probes.

As mentioned earlier, SHIVA comprises two sensor heads, one viewing along the nominal Parker spiral interplanetary magnetic field lines and the other perpendicular to the field lines. The instrument has a configurable onboard processing to set the cadence and energy resolution.

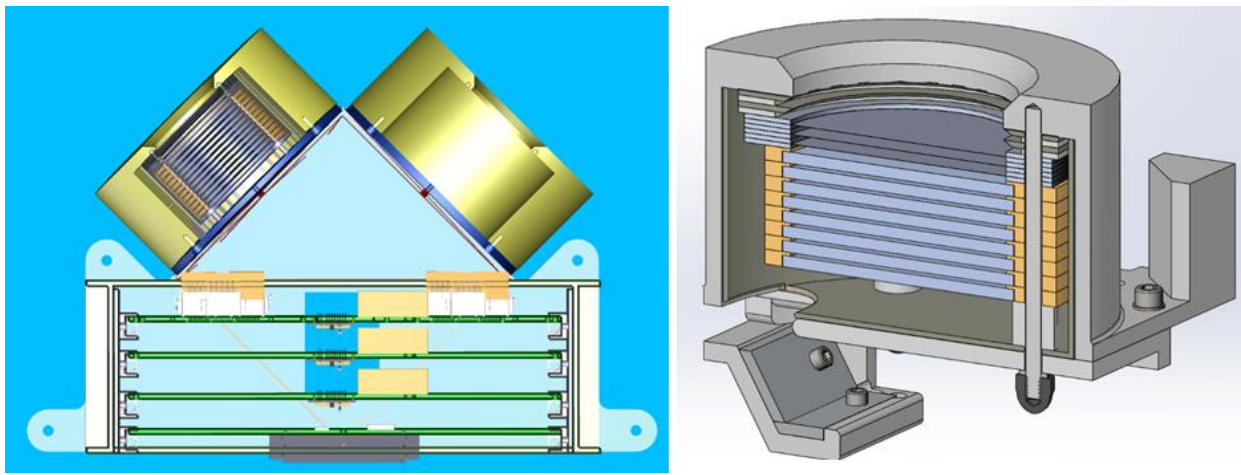

**Figure 12**. Schematic of the SHIVA instrument with the two sensor heads at the top (left) and the cross-sectional view of one of the sensors (right). The geometry factor is 31 $cm^2$ sr (SSD) and 0.05 $cm^2$ sr (APD). A cut-away rendering of the sensor head is shown to the right with each detector stack surrounded by an inner tungsten and an outer aluminum shielding (APDs are not shown). The electronics box below the sensor heads comprises front end electronics cards and the on-board processor.

### 3.12 Payload Accommodation

The spacecraft MOST1&2 with the ten scientific instruments each, the high-gain antenna (HGA), and the solar panels are shown in Figure 13 in stowed and deployed configurations. The spacecraft design used for the Earth Affecting Solar Causes Observatory (EASCO) mission (Gopalswamy et al., 2011a,b) has been adapted for MOST. The spacecraft bus will be a rectangular composite honeycomb structure, with a 62-inch separation system. The spacecraft are three-axis-stabilized. The cluster of remote sensing telescopes (MaDI, ICIE, HXI, and WCOR) is placed together on the Sun-facing side of the spacecraft and are actively pointed to. The HIP instrument is mounted on a platform to clear the HGA. FETCH's log-periodic dipole antenna will be mounted on a boom to prevent any light scattered from the antenna entering into WCOR aperture.

MOST3&4 carry only the FETCH equipment, so the spacecraft bus is very simple. Figure 14 shows the stowed and deployed configurations. The log-periodic dipole antenna is attached directly to the spacecraft.

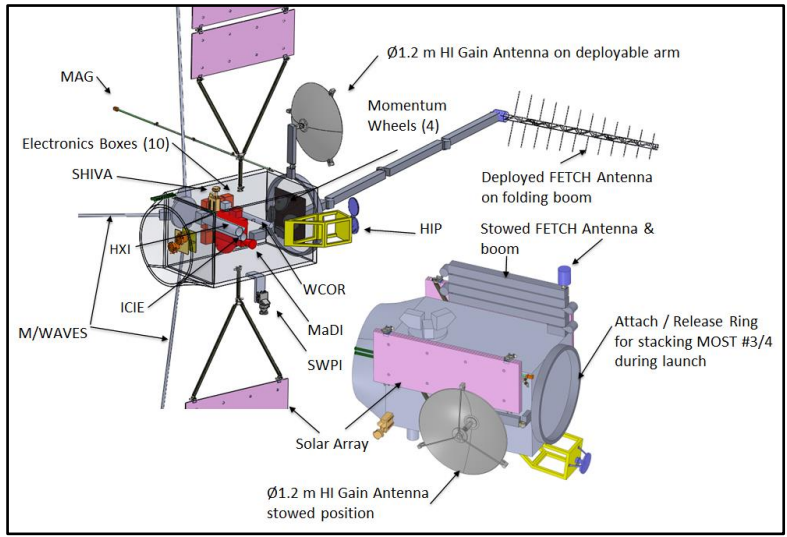

**Figure 13**. MOST1&2 with the full instrument suite showing stowed and deployed configurations. All the instruments are marked. The high-gain antenna points to Earth. The remote-sensing instruments except HIP and FETCH point to the Sun. The MAGs will be mounted on the MAG boom shown. The FETCH antenna is ~3.3 m long and the longest dipole is ~1 m long and fits into the canister (0.25 m diameter and 0.3 m long) shown in purple.

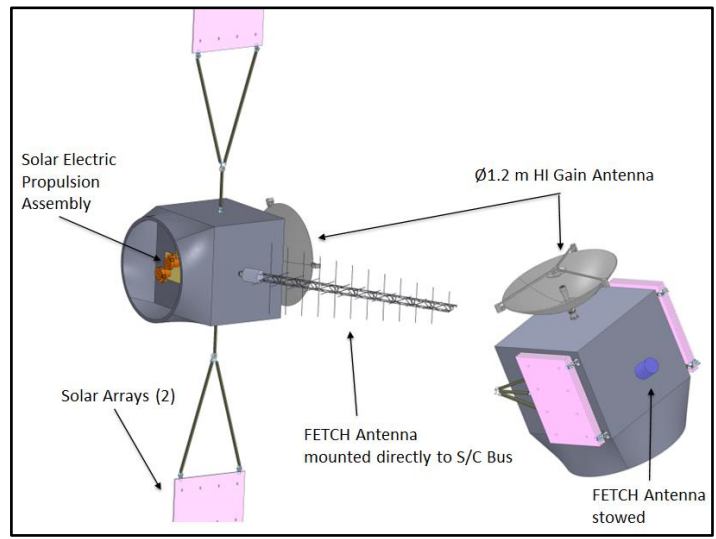

**Figure 14**. MOST3&4 with the FETCH instrument shown in deployed and stowed configurations. The solar electric propulsion assembly and the deployed FETCH antenna are pointed to in the deployed spacecraft. The purple cylinder in the stowed FETCH antenna. The FETCH antenna is mounted directly on the spacecraft (no boom).

The stowed MOST spacecraft are mounted in a Vulcan dual manifest fairing in two groups as shown in Figure 15. The upper pair consists of MOST1&3, while MOST2&4 are in the lower pair. We considered several options for the launch vehicle: Antares, Vulcan, Falcon Heavy, Falcon 9 (Full Thrust, RTLS/ASDS). Their performances are shown in Figure 15 (right) in terms of mass delivered as a function of characteristic energy, c3. Antares and the Return To Launch/Landing Site (RTLS) variant of Falcon 9 (Full Thrust) cannot meet our mass requirements, the other three can deliver the estimated mass of ~3000 kg for MOST. Of these, we selected Vulcan because of its dual payload fairing suitable for launching all the MOST spacecraft together (see Fig. 15, left).

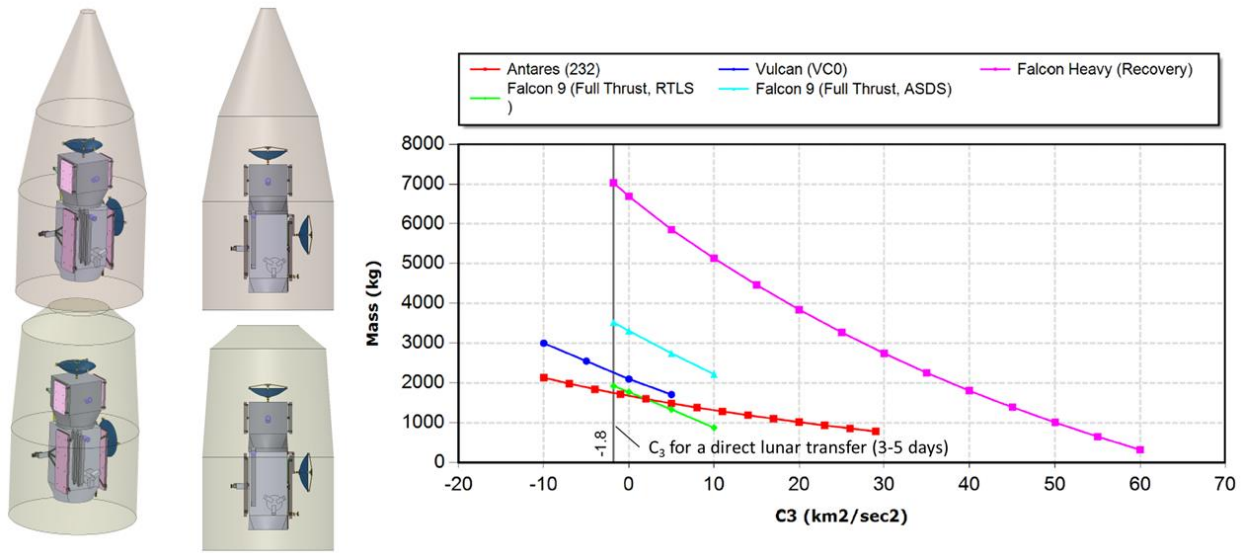

**Figure 15**. (left and middle) Two views of the MOST launch configuration in Vulcan's split manifest fairing. MOST3 is on top of MOST1 in the upper pair; MOST4 is on top of MOST2 in the lower pair. (right) Characteristic energy (c3) vs. mass that can be delivered for several launch vehicles. The vertical line at $c3 = -1.8$ km$^2$s$^{-2}$ corresponds to the direct lunar transfer orbit. We selected the Vulcan based on the preliminary estimate of mass (~3000 kg) and the need to launch the 4 MOST SC together enabled by Vulcan's large, dual payload fairing to hold two sets of two stacked spacecraft.

### 3.13 Flight Dynamics and Orbital Selection

The flight dynamics (FD) analysis started with the requirement that MOST1 and MOST2 should be parked at L4 and L5, respectively, while MOST3 and MOST4 drift beyond L4 and L5, to a maximum Earth-Sun-spacecraft angle (ESSA) of 80⁰. The locations of MOST3 and MOST4 are denoted by L4' and L5', respectively to indicate these locations change over the mission lifetime including a one-year sit and stare at ESSA ~80⁰. The analysis considered two cases that differ in their arrival times at L4 and L5 by ~1 year. The spacecraft are placed in the desired locations by performing lunar flyby similar to how the STEREO spacecraft were placed into their desired heliocentric drift-away orbits. The main difference is that MOST1 and MOST2 are stopped for a sit-and-stare from L4 and L5, respectively. MOST3 and MOST4 are drifting but bounded by the maximum ESSA of ~80⁰. This study assumed a wet mass of 600 kg for MOST1&2 and 400 kg for MOST3&4. The analysis considered two types of propulsion: high-thrust (chemical) and low-thrust (solar electric propulsion).

From FD point of view, there are four phases to the MOST mission: 1. launch and lunar flyby, 2. transfer phase toward the desired locations (L4, L4') and (L5, L5'), 3. dwell phase when all spacecraft are in place for a steady one year of observations, and 4. drift phase when MOST3&4 drift toward Earth and, at mission's end, each of these two spacecraft occupies the original position of the other. In the launch and lunar flyby phase, two initial conditions were considered, the difference being the arrival times at the final points by ~1 year. For this study, chemical propulsion was used for the lunar flyby phase, but the same can be accomplished using a low-thrust electric propulsion. In the drift phase, the drift back towards Earth was modeled with an impulsive burn but could be modeled with a low-thrust architecture as well. All four spacecraft are launched together into a direct lunar transfer orbit (5-day transfer period) as illustrated in Figure 16. Each spacecraft will perform a trajectory correction maneuver (TCM) in order to plan a particular lunar flyby. The flybys place the upper and lower constellations into heliocentric drift-away orbits: towards the L4 point (upper constellation) and toward L5 point (lower constellation). The constellation $\Delta V$ for various maneuvers along with the time since launch ($\Delta T$ in units of days, d and years, y) are listed in Table 6 for a drift of 15º per year between the upper and lower constellations, the initial configuration is established in 5.72 years. For MOST1-4, the $\Delta V$ requirements for the full mission are ~464, ~494, ~522, and ~521 m/s, respectively. When the initial configuration is desired to be established a year early, the $\Delta V$ requirements nearly double for each spacecraft.

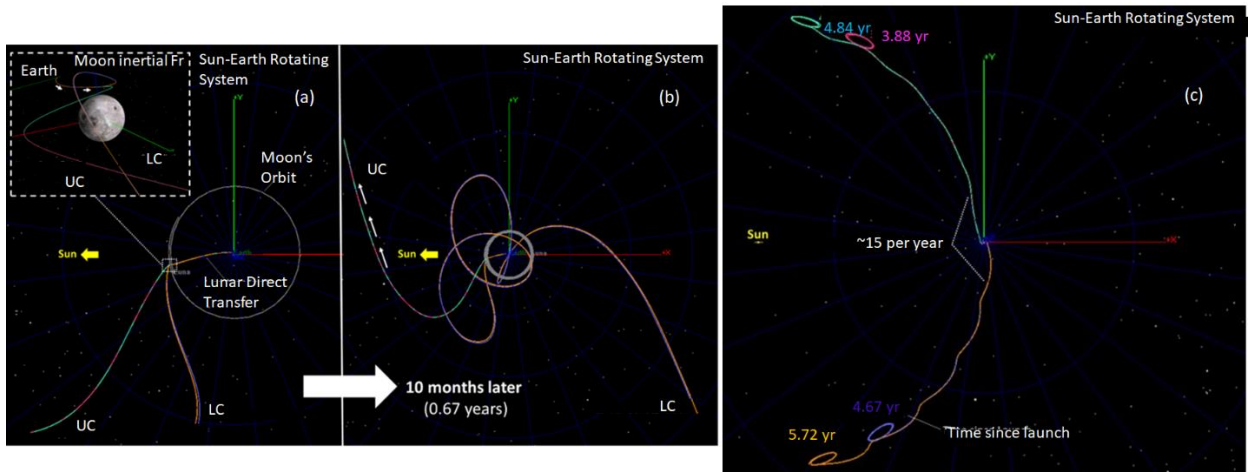

**Figure 16**. Results of a flight dynamics analysis for a separation angle of 15º per year between the upper (UC) and lower (LC) constellations. (a) early after launch, (b) 10 months into the mission, and (c) when the initial configuration of all the spacecraft is established. MOST1-4 arrive in their respective locations at 3.88, 4.67, 4.84, and 5.72 years, respectively (see Table 6).

The impulsive maneuver necessary for MOST3&4 to turn around from the dwell location consists of a series of two maneuvers: the first one adjusts the semi-major axis of the orbit to induce the necessary drift rate; the second one circularizes the orbit to stabilize the drift rate with respect to the constellation and Earth. MOST3 and MOST4 will exchange their initial positions after ~9.5 years with a $\Delta V$ cost of 496 m/s (MOST3) and 499 m/s (MOST4). The turnback and drift achieved by the impulsive maneuvers can also be accomplished by a low-thrust architecture.

**Table 6.** Constellation $\Delta V$ budget

|  |  | MOST1 | MOST2 | MOST3 | MOST4 |
| --- | --- | --- | --- | --- | --- |

| Maneuver Number | Maneuver Purpose | ΔT | ΔV m/s | ΔT | ΔV m/s | ΔT | ΔV m/s | ΔT | ΔV m/s |
|---|---|---|---|---|---|---|---|---|---|
| 1 | Target Ideal Lunar Flyby | 1d | 6.9 | 1d | 32.0 | 1d | 6.4 | 1d | 32.0 |
| 2 | Powered Flyby | - | - | 3.5d | 5.9 | - | - | 3.5d | 5.9 |
| 3 | Insertion (Nullify Drift) | 3.88y | 456.8 | 4.67y | 456.5 | 4.84y | 515.9 | 5.72y | 482.8 |
| **Total** | | | **463.7** | | **494.4** | | **522.3** | | **520.7** |

## 3.14 MOST Project Life Cycle

Figure 17 shows the project life cycle of the MOST mission including the notional phases. It takes about 8 months for phase A studies (preliminary analysis and mission definition), and 11 months for phase B (system definition, preliminary design, and review), one year of phase C (final design and critical design review), 26 months of phase D-1 (subsystem development, spacecraft integration and testing), and 1 month of phase D-2 (launch and check out), and 10 years for phase E/F (science operations). During E and F, the constellations will drift toward L4 and L5, taking about 5 years; all the spacecraft will be in the dwell phase for a year followed by MOST3 and MOST4 drifting toward Earth, while MOST1 and MOST2 are places in halo orbits around L4 and L5, respectively. In the extended mission, MOST3 and MOST4 will occupy each other's dwell position after another 5 years (after ~11 years from launch). All instruments will start scientific operations in the cruise phase (enroute to L4 ad L5). Reclosable doors will be opened and closed as needed. Table 7 summarizes the crude cost estimates of the hardware, wraps (management, system engineering), and launch vehicle, indicating that MOST is a large mission with a total cost of ~$900 M. MOST will be a Great Observatory with a cost less than half the cost of the BepiColombo mission. Given the large swath of the heliophysics community that will use the data, the benefit far outweighs the cost.

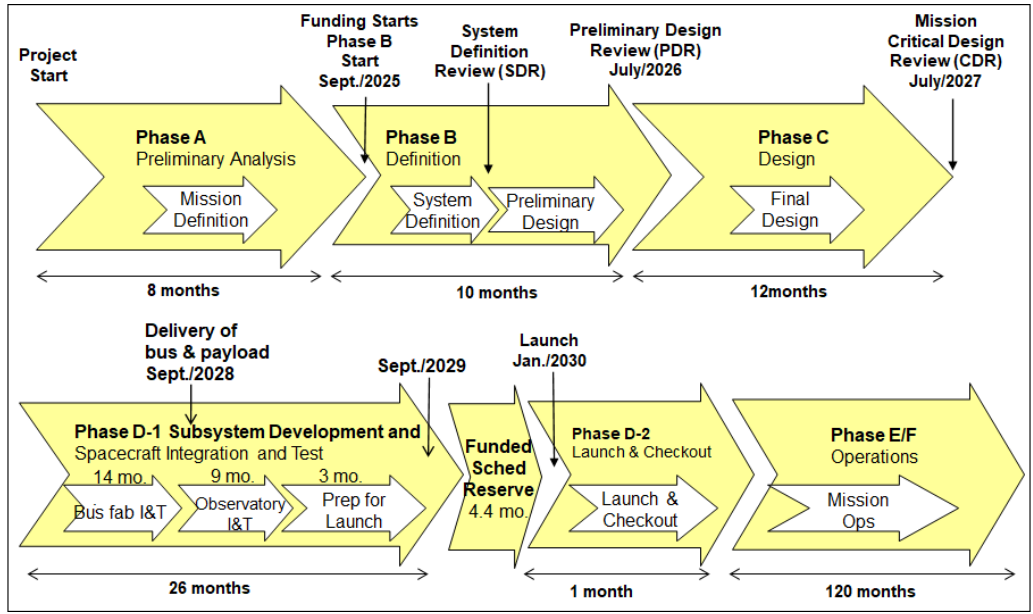

**Figure 17**. Project life cycle of the MOST mission, showing the extents of various phases and the tasks to be completed.

Table 7. Estimated Mission Cost

| Item | Description | Cost (M$) | Remark |
| --- | --- | --- | --- |
| Instrument cost | 9 instruments × 2, FETCH × 4 | 128 | All but FETCH: EASCO heritage |
| MOST1&2 | 2 buses | 168 | 150×1.12, EASCO |
| MOST3&4 | 2 buses | 84 | 75×1.12, EASCO |
| WRAPS | Instruments + S/C | 351 | 380 × 0.925 |
| Total cost | Instruments+S/C+WRAPS | 731 | If launch provided |
| Launch vehicle (LV) | Vulcan with dual manifest fairing | 175 | Rough estimate |
| Total cost | Mission+WRAPS+LV | 906 | <$1B |

## 4 Payload Synergy and Modeling

### 4.1 Synergy among MOST instruments

The instrument suites provide imagery and time series data to reveal magnetic connectivity across solar and heliospheric domains. Data from different combinations of multiple, complementary MOST instruments are needed to bring closure on major mission-wide, global-scale science objectives. Table 8 summarizes the specifications of the MOST instruments, which indicate that the data coverage is quite extensive and different combinations can be used for different investigations.

For example, to achieve objective 1.2 (Determine the complete life cycle of active regions) we need to use a combination of data from MaDI, ICIE, and HXI to characterize the complete lifecycle of active regions from emergence to dissipation. It has been shown that the farside seismic signatures are correlated with the nearside magnetic signatures (Gonzalez Hernandez et al., 2007) observed when the active regions are on the nearside, in spite of the expected evolution over a period of 1-2 weeks before they reach farside. MOST can greatly improve the situation because farside imaging becomes possible from three views: E60 (L5), W00, and W60 (L4) instead of just W00 (Earth view). In addition, the magnetograms at L4 and L5 will greatly reduce the time between the magnetogram and farside helioseismic observations to just a couple of days. Identification and assigning magnetic structure from such combinations will greatly improve solar wind modeling and solar irradiance forecasting (Fontenla et al., 2009).

To characterize the global coronal magnetic connectivity from solar surface to the heliosphere and its slow evolution over timescales ranging from the solar rotation to solar cycle we need data from MaDI, WCOR, HIP, FETCH, MAG, and SWPI. In order to characterize the origin and energetics of solar eruptions as they propagate to the heliosphere, we need to use data from all ten MOST instruments.

The FETCH instrument is unique in that it provides detailed information on the magnetic field at each sampled point in contrast to imaging methods, which provide mostly distributions of material density across space. The synergy between FETCH and HIP provides an opportunity to constrain the density and line-of-sight magnetic field.

**Table 8**. High-level specifications of the science instruments

| Instrument | FOV | Spatial resolution | Temporal Resolution | Mass (kg) | Average Power (W) | Data Rate (kbps) |
|---|---|---|---|---|---|---|
| MaDI | Full disk | 2" | 90 min[a], 1 min[b] | 6 | 20 | 140 |
| ICIE | 0 - 3 Rs | 2" | 1 min | 30 | 35 | 48 |
| HXI | Full Disk | 7-100" | 0.1 -1 s | 6.5 | 8 | 1 |
| WCOR | 2.5 – 15 Rs | 30" | 1-15 min | 20 | 20 | 15 |
| HIP | 10-220 Rs | 2'-4' | 20 min | 16 | 20 | 15 |
| FETCH | 18-107 Rs | -- | 100 s | 88 | 515 | 2 |
| M/WAVES | 2 –215 Rs | -- | 1 min | 13 | 15 | 2 |
| SWPI | In situ | -- | 1 min | 5 | 7.7 | 16.8 |
| MAG | In situ | -- | 1 min | 1.2 | 1 | 2 |
| SHIVA | In situ | -- | 1 min | 4 | 5.5 | 30 |

[a]Magnetograms, [b]Dopplergrams

It is worth noting the possibility of synergy between MOST and ESA's Vigil mission ((Kraft et al. 2017). The Vigil mission uses a subset of instruments proposed for the L5 EASCO mission concept a decade ago (Gopalswamy et al. 2011a,b) and further discussed during L5 consortium meetings (https://cdaw.gsfc.nasa.gov/meetings/2019_L5C/). The subset was selected based on the Vigil mission's focus on nowcasting and forecasting. MOST will carry the full suite of instruments proposed in EASCO and placed not only at L5, but also at L4. Two additional spacecraft will be used exclusively for FETCH observations in combination with radio equipment at L4 and L5. Thus, MOST is focused on science, providing the imagery and time series data needed for wide ranging heliophysics investigations. If MOST and Vigil have overlapping period of operations, one expects good synergy between them. MOST can complement Vigil observations by providing low frequency data on Earth-approaching interplanetary shocks at various distances between the Sun and 1 AU (M/WAVES). The shocks are also responsible for producing SEP events observed by SHIVA. MOST will also provide information on flare imagery for enhancing Vigil's space weather forecasting. Vigil will measure photospheric magnetic field; MOST will measure photospheric and chromospheric fields (MaDI), and the magnetic field in the inner heliosphere (FETCH). FETCH's magnetic field estimates can aid Vigil's space weather forecasting by providing magnetic field estimates in CMEs from FETCH and 3-D modeling using the FRED (Flux Rope from Eruption Data) technique (Gopalswamy et al. 2018). Vigil's observations will be useful in cross calibration with the three common instruments: coronagraph, heliospheric imager, and magnetograph (with a possible addition of an EUV instrument from NASA).

## 4.2 Modeling

Numerical modeling and simulations are essential in achieving the objectives of the MOST mission. Models of the background solar wind and the transients propagating in the solar wind are actively pursued (see e.g., Van der Holst et al., 2014; Jin et al. 2017a,b; Manchester et al., Sachdeva et al.; Odstrcil et al., 2020) because they provide the global context needed for a better understanding of the Sun-heliospheric system. As noted, the photospheric magnetic field is a key driver of the solar wind models. In principle, the corona and solar wind models need the instantaneous magnetic field distribution over the entire solar surface including the poles. Although MOST will not cover the whole $4\pi$ steradians, the combination of photospheric magnetograms from L1, L4 and L5 is very close to that ideal objective (Pevtsov et al., 2020).

The STEREO mission demonstrated the importance of multiple views of solar features such as prominences, streamers, and CMEs. As major players in the Sun-Earth system variability, CMEs need to be characterized as early as possible, especially in the coronagraph FOV. CME modeling will help with both interpreting the WCOR observations and using them to simulate CME behavior at farther distances. This first part is a critical, but often overlooked aspect of CME modeling. Coronagraphs, like most remote-sensing instruments observing an optically thin plasma, integrate along the line of sight, compressing 3-D information into a 2-D plane. Therefore, some form of geometric modeling is needed to extract maximum information on the observed phenomena. Modeling is also needed for instruments observing optically thick photospheric plasma. For example, MaDI measurements need to be modeled to obtain the magnetic field in the photosphere. Similarly, helioseismological inversion techniques need to be employed to obtain physical properties in the convection zone. In the case of coronagraph observations, a cone model (e.g., Fisher and Munro, 1984; Zhao et al., 2002; Na et al., 2013, 2017) or the graduated cylindrical shell model (GCS; Thernisien et al., 2006, 2009) need to be used.. If one attempts to reconstruct a CME using a single viewpoint there is often a degeneracy of plausible CME parameters that lead to "suitable" visual agreement between a wireframe model and the coronagraph image. Figure 18 shows three different fits to the same synthetic coronagraph image (from Verbeke et al., 2022). Despite appearing nearly the same visually, the CME parameters corresponding to these reconstructions vary by 34° in angular width and 5° in latitude as noted on the plots. When the number of viewpoints increases, the uncertainty in the derived parameters is reduced significantly.

These authors demonstrate the need for at least two viewpoints to reduce the uncertainty in the derived CME parameters.

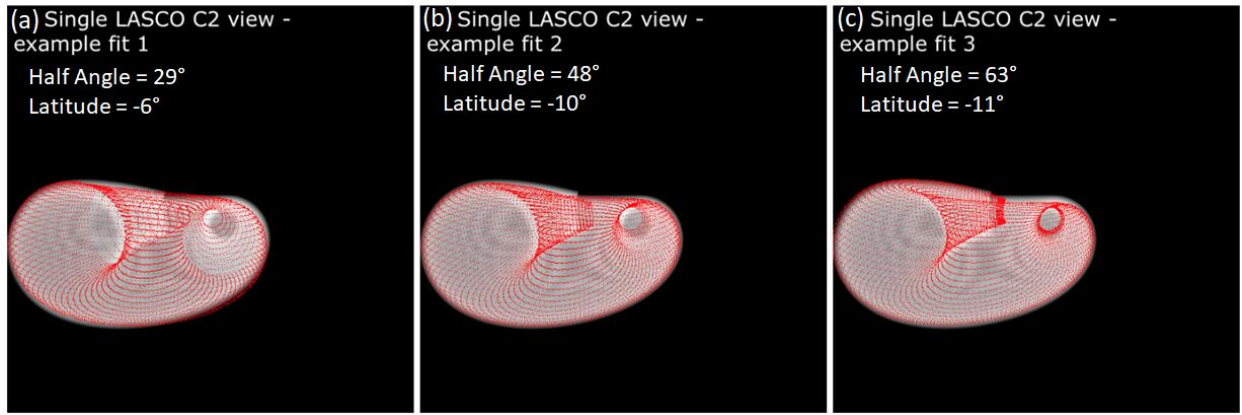

**Figure 18**. Three different fits to the same synthetic coronagraph image. Despite appearing nearly the same visually, the CME parameters such as flux rope width and latitude corresponding to these reconstructions vary significantly as shown on the plots (from Verbeke et al., 2022).

Global heliospheric models solving time-dependent MHD equations are popular and powerful tools to investigate the propagation, evolution, and space weather potential of CMEs throughout interplanetary space (e.g. Mays et al., 2015; Török et al., 2018; Asvestari et al., 2021; Scolini et al., 2019, 2020). These models, particularly those designed for space weather research and forecasting purposes, typically initiate CMEs near 0.1 au (21.5 Rs), i.e., beyond the Alfvén point, to limit computational costs while retaining a realistic description of plasma structures in the solar wind (e.g., Odstrcil, 2003; Shiota and Kataoka, 2016; Pomoell and Poedts, 2018). Performing accurate estimations of the complete set of CME initial parameters near 0.1 au is critical for a realistic modeling of CME propagation through the interplanetary space (Scolini et al., 2019, 2020; Singh et al., 2019). Geometric reconstruction of CME flux ropes is routinely done using EUV, coronagraph, and heliospheric imaging. The flux rope's magnetic properties can be derived from the FRED technique, which assigns the total reconnected flux derived from the photospheric magnetogram and EUV images of the PEA (Gopalswamy et al. 2018). The derived flux rope parameters can be converted to the inputs required by these heliospheric models, either by assuming no change between 15-21.5 Rs, or by assuming some sort of scaling with distance. Figure 19 shows the synergy among various MOST instruments. While MaDI, ICIE, and WCOR contribute to key input parameters to the global MHD models, HIP and FETCH provide key constraints in validating the models. As summarized in Figure 19, WCOR observations will enable routine determination of the morphology, kinematics, geometry, and thermodynamics of CMEs in the middle-to-upper corona and will be pivotal to the interpretation of low coronal (i.e., from MaDI and ICIE) and heliospheric (i.e., from HIP and FETCH) observations, allowing investigations of CME magnetic structures through stereoscopic observations obtained across various heliocentric distances.

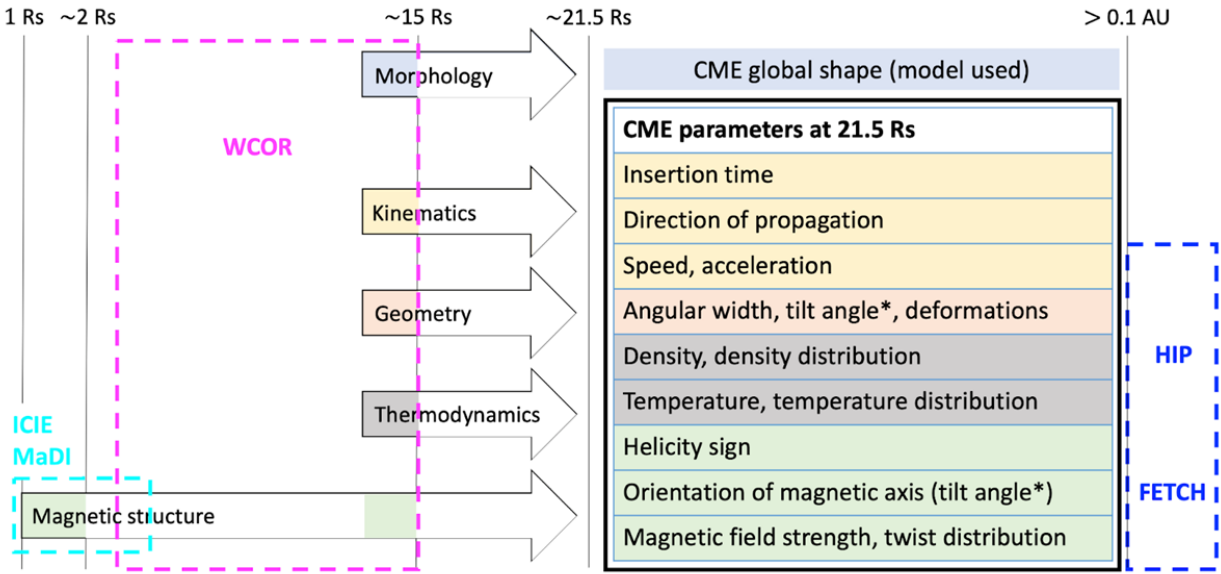

**Figure 19**. Schematics of the input parameters typically required by heliospheric MHD models initiating CMEs near 0.1 au (21.5 Rs), including contributions from WCOR observations, and possible synergies with other MOST instruments.

## 5   Mission Operations

We have not discussed standard items like power system and avionics. The power system will be designed with the solar panel taking into account of normal observatory operations and solar electric propulsion. MOST will use star trackers (roll knowledge) and guide telescopes (pointing accuracy) similar to the ones on STEREO. The MOST avionics includes (i) Integrated Avionics Unit: uses typical command and data handling cards; handles data storage using solid-state recorders; manages attitude control, and spacecraft power/battery management. (ii) Redundancy Management Unit: manages primary and redundant power interfaces to Battery/Solar Array. (iii) Gimbal Control Electronics: controls dual-axis gimbals for propulsion and antennas. The flight hardware/software is based on proven in-house or commercial off-the-shelf (COTS) system and no credible technical risk has been identified. Radiation analysis will be performed in the future to minimize the radiation dose to acceptable level using aluminum shielding. Also to be done in the future is to validate all designs with appropriate reliability analyses – Fault Tree Analysis, Failure Mode and Effects Analysis, Parts Stress Analysis, Probabilistic Risk Analysis, and Worst Case Analysis.

The Mission Operation Center (MOC) and the Science Operation Center (SOC) of the MOST mission will be located at NASA/GSFC. The MOC will handle the following functions: mission planning and scheduling, orbit determination/control, network and contact scheduling, commanding, spacecraft monitor/control, real-time health/safety processing, trending/analysis, instrument data handling, level 0 product processing, and level 0 data archive. The MOC implementation will use existing tools and software. There are no mission requirements that drive technology; technology required is readily available and operational today for several spacecraft. The MOC will also handle infrequent calibration rolls, momentum dumps, close instrument doors where needed, deployments (solar array, MAG boom, and FETCH antennas), and orbital maneuvers (solar electrical propulsion thrusts).

DSN will be used to cover all the critical events of the mission: separation from launch vehicle, attitude acquisition, solar array deployment, and propulsion system tests. If the separation is not in the view of the ground station, a portable ground station will be used. The mission operation plan includes five elements: (i) Nominal Sequence Planning and Commanding: receive instrument commands from SOC five days per week; uplink command sequences every weekday from the MOC. (ii) Operations staffing: 8 hours per day, five days per week operations by MOC staff; autonomous monitoring when unstaffed; designated operations team members will be alerted in the event of a problem or opportunity. (iii) Operations Training: operations team will participate in spacecraft integration and testing; also performs mission simulations prior to launch to verify readiness. (v) Operations Center Development: Reuses existing facility and software.

# 6     Summary and Conclusions

We presented the MOST mission concept that will build upon the successes of SOHO and STEREO missions with multiple new views of the Sun and enhanced instrument capabilities. The MOST mission is envisioned as the next generation Great Observatory to provide necessary imagery and time-series data of the Sun and heliosphere to understand the magnetic coupling between the solar interior and the extended atmosphere. The MOST mission is focused on understanding the global impact of flux emergence from the solar interior – from the inner corona out to 1 au. MOST is a multi-spacecraft mission in Earth orbit around the Sun positioned to obtain three-dimensional information on large-scale heliospheric structures such as coronal mass ejections, stream interaction regions, and the solar wind itself. MOST will consist of two pairs of spacecraft located in the vicinity of L4 and L5. The spacecraft stationed at L4 and L5 will carry seven remote-sensing and three in-situ instrument suites. MOST will also carry a novel radio package FETCH carrying transmitters and receivers on all four spacecraft to measure the magnetic content of solar wind structures using the Faraday rotation technique.  The MOST mission will be able to sample the magnetized plasma between the Sun and Earth during the mission lifetime.  It is expected that MOST will be a significant part of the next generation Heliophysics System Observatory benefiting a large swath of the heliophysics community.

The main conclusions of this study can be summarized as follows.

1. Only a Great Observatory with an optimal set of remote-sensing and in-situ instruments can provide all the imagery and time-series data needed for system science.

2. The Sun-Earth system variability is driven by solar magnetism, so it is necessary to measure the magnetic field in the photosphere, chromosphere, corona, and the interplanetary medium leading to breakthroughs on critical questions.

3. FETCH is a novel concept requiring the analysis of spacecraft-to-spacecraft radio signals to provide magnetic field measurements from the outer corona to 1 au.

4. Most of the instruments have high heritage and TRL >6, except FETCH, which needs further study to optimize signal-to-noise ratio for FR measurements and to minimize the mass and power. MADI based on traditional magnetograph concept (e.g., CDM) is also at TRL >6, but magnetograph based on the revolutionary IPSOS concept requires further study.

5. The instrument FOVs are optimized to provide a continuous spatial coverage from the Sun to 1 au.

6. The mechanical assembly of the instruments on the spacecraft closely follows the STEREO mission, except for the boom requirement for FETCH.

7. The launch vehicle appropriate to the MOST mission has been found to be Vulcan with a split manifest fairing. MOST1&3 and MOST2&4 will be paired in the launch configuration.

8. Flight dynamics studies indicate that electric propulsion is a viable option. More trade studies will be performed between chemical and electric propulsions.

9. The prime mission has a duration of ~11 years (cruise, dwell, and drift). The extended mission will prolong the mission for another five years when MOST3 and MOST4 will switch their dwell positions.

10. MOST will be a large mission costing about $900 M.

# 7 Conflict of Interest

The authors have no conflict of interest to declare.

# 8 Author Contributions

NG contributed to conception and design of the mission study and wrote the first draft of the manuscript. NH and AP organized the MaDI group and provided instrument design. LG contributed the design of ICIE. SK provided the HXI design. PN, QG, JZ, and NG contributed to the WCOR design. CD contributed to the HIP instrument design. LJ, SF, LL, and NG contributed to the development of the FETCH concept. SB and NG contributed to the update of the WAVES instrument. MD and PM developed the SWPI design. JG provided the MAG design. SK developed SHIVA. WM, CK, and CS contributed to the modeling section. All authors contributed to manuscript revision, read, and approved the submitted version.

# 9 Funding

The MOST concept study was funded by NASA Goddard Space Flight Center's Heliophysics Line of Business (LOB), the Internal Research and Development (IRAD) program, and the STEREO project.

# 10 Acknowledgments

The team thanks K. Parsay, L. Purves, G. Voellmer, M. Deshpande, and M. Shelton for engineering support. NG thanks the NASA Goddard Space Flight Center's Heliophysics Line of Business (LOB), IRAD program, and the STEREO project for support. The National Solar Observatory (NSO) is operated by the Association of Universities for Research in Astronomy,

Inc. (AURA), under cooperative agreement with the National Science Foundation. K-SC is supported by the KASI Qrontier (Promising for the future) L4 project.

**References**

Amerstorfer, T., Hinterreiter, J., Reiss, M.A., Möstl, C., Davies, J.A., Bailey, R.L. et al. (2021). CME arrival prediction using ensemble modeling based on heliospheric imaging observations. *Space Weather* 19, e2020SW002553, doi: 10.1029/2020SW002553

Asvestari, E., Pomoell, J., Kilpua, E., Good, S., Chatzistergos, T., Temmer, M., et al. (2021). Modelling a multi-spacecraft coronal mass ejection encounter with EUHFORIA. *Astron. Astrophys.* 652, iA27. doi:10.1051/0004-6361/202140315

Bale, S. D. Ullrich, R., Goetz, K. Alster, N. Cecconi, B. Dekkali, M., et al. (2008). The Electric Antennas for the STEREO/WAVES Experiment. *Space Sci. Rev.* 136, 529–547. Doi:10.1007/s11214-007-9251-x

Bale, S. D., Goetz, K., Harvey, P. R., Turin, P., Bonnell, J. W., Dudok de Wit, T., et al. (2016). The FIELDS Instrument Suite for Solar Probe Plus. Measuring the Coronal Plasma and Magnetic Field, Plasma Waves and Turbulence, and Radio Signatures of Solar Transients. *Space Sci. Rev.* 204, 49-82. doi:10.1007/s11214-016-0244-5

Bale, S. D., Badman, S. T., Bonnell, J. W., Bowen, T. A., Burgess, D., Case, A. W., et al. (2019). Highly structured slow solar wind emerging from an equatorial coronal hole. *Nature* 576, 237–242. doi:10.1038/s41586-019-1818-7

Bale, S. D., Horbury, T.S., Velli, M., Desai, M. I., Halekas, J. S., McManus, M. D., et al. (2021). A Solar Source of Alfvenic Magnetic Field Switchbacks: In Situ Remnants of Magnetic Funnels on Supergranulation Scales. *Astrophys. J.* 923, 174. doi:10.3847/1538-4357/ac2d8c

Bemporad, A. (2021). Possible Advantages of a Twin Spacecraft Heliospheric Mission at the Sun-Earth Lagrangian Points L4 and L5. *Front. Astron. Space Sci.* 8, 627576, doi: 10.3389/fspas.2021.627576

Berghmans, D., Hochedez, J. F., Defise, J. M., Lecat, J. H., Nicula, B., Slemzin, V., et al. (2006). SWAP onboard PROBA 2, a new EUV imager for solar monitoring. *Adv. Space Res.* 38, 1807–1811. doi:10.1016/j.asr.2005.03.070

Berghmans, D., Antolin, P., Auchère, F., Aznar Cuadrado, R.. Barczynski, K., Chitta, L. P. et al. (2023). First perihelion of EUI on the Solar Orbiter mission. *A&A* 675, A110 (2023); doi:10.1051/0004-6361/202245586

Bernfeld M., Cook C. E., Paolillo J., and Palmieri C. A., (1965). Matched Filtering, Pulse Compression and Waveform Design. *Microwave Journal*, Oct 1964 – Jan 1965 (four issues).

Bird, M. K. (2007). Coronal Faraday rotation of occulted radio signals. *Astron. Astrophys. Trans.* 26, 441–453. Doi:10.1080/10556790701595236

Bougeret, J.-L., Goetz K., Kaiser, M. L., Bale, S. D., Kellogg, P. J., Maksimovic, M., et al. (2008). S/WAVES: The Radio and Plasma Wave Investigation on the STEREO Mission. *Space Sci. Rev.* 136: 487–528. doi:10.1007/s11214-007-9298-8

Brueckner, G. E., Howard, R. A., Koomen, M. J., Korendyke, C. M., Michels, D. J., Moses, J. D., et al. (1995). The Large Angle Spectroscopic Coronagraph (LASCO). *Solar Phys.* 162, 357–402. doi:10.1007/BF00733434

Burch, J. L., Goldstein, R., Cravens, T. E., Gibson, W. C., Lundin, R. N., Pollock, C. J. et al. (2006). RPC-IES: The Ion and Electron Sensor of the Rosetta Plasma Consortium. *Space Sci. Rev.* 128, 697–712. doi: 10.1007/s11214-006-9002-4

Cane, H. V., Sheeley, N. R., and Howard, R. A. (1987). Energetic interplanetary shocks, radio emission, and coronal mass ejections. *J. Geophys. Res.*, 92, Issue A9, 9869, doi: 10.1029/JA092iA09p09869

Christensen-Dalsgaard, J. (2021). Solar structure and evolution. *Living Rev Sol Phys* 18, 2, doi: 10.1007/s41116-020-00028-3

Collett, E. (1992). Polarized Light: Fundamentals and Applications. CRC Press, Boca Raton, FL.

Cummer, S. A., Reiner, M. J., Reinisch, B. W., Kaiser, M. L., Green, J. L., Benson, R. F., et al. (2001). A test of magnetospheric radio tomographic imaging with IMAGE and WIND. *Geophys. Res. Lett.* 28, 1131–1134. doi:10.1029/2000GL012683

Cummer, S. A., Green, J. L., Reinisch, B. W., Fung, S. F., Kaiser, M. L., Pickett, J. S., et al. (2003). Advances in magnetospheric radio wave analysis and tomography. *Adv. Space Res.* 32, 329–336. doi:10.1016/S0273-1177(03)90271-7

DeForest, C. E., Killough, R., Gibson, S. E., Henry, A. M. Case, T., Beasley, M., Laurent, G., et al. (2022). Polarimeter to UNify the Corona and Heliosphere (PUNCH): Science, Status, and Path to Flight. 2022 IEEE Aerospace Conference (AERO). 1–11. doi:10.1109/AERO53065.2022.9843340

Delaboudinière, J.-P., Artzner, G. E., Brunaud, J., Gabriel, A. H., Hochedez, J. F., Millier, F., et al. (1995). EIT: Extreme-Ultraviolet Imaging Telescope for the SOHO Mission. *Solar Phys.* 162, 291–312. doi:10.1007/BF00733432

Dere, K. P.; Brueckner, G. E.; Howard, R. A., Koomen, M. J., Korendyke, C. M., Kreplin, R. W., et al. (1997). EIT and LASCO Observations of the Initiation of a Coronal Mass Ejection. *Solar Phys.* 175, 601–621. , doi:10.1023/A:1004907307376

Domingo, V., Fleck, B., and Poland, A. I. (1995). The SOHO Mission: an Overview. *Solar Phys.* 162, 1–37. doi:10.1007/BF00733425

Eyles, C. J., Harrison, R. A., Davis, C. J., Waltham, N. R., Shaughnessy, B. M., Mapson-Menard, H. C. A., et al. (2009). The Heliospheric Imagers Onboard the STEREO Mission. *Solar Phys.* 254, 387–445. doi:10.1007/s11207-008-9299-0

Fisher, R. R., and Munro, R. H. (1984). Coronal transient geometry. I-The flare-associated event of 1981 March 25. *Astrophys. J.* 280, 428–439. doi:10.1086/162009

Fontenla, J. M., Quémerais, E., González Hernández, I., Lindsey, C., Haberreiter, M. (2009). Solar irradiance forecast and far-side imaging. *Adv. Space Res.* 44, 457–464. doi:10.1016/j.asr.2009.04.010

Fox, N. J., Velli, M. C., Bale, S. D., Decker, R., Driesman, A., Howard, R. A. et al. (2016). The Solar Probe Plus Mission: Humanity's First Visit to Our Star. Space Sci. Rev. 204, 7, doi: 10.1007/s11214-015-0211-6

Golub, L., DeLuca, E., Austin, G., Bookbinder, J., Caldwell, D., Cheimets, P., et al. (2007). The X-Ray Telescope (XRT) for the Hinode Mission. *Solar Phys.* 243, 63–86. doi:10.1007/s11207-007-0182-1

Golub, L., and Savcheva, S. L. (2016). COSIE: A Wide-field EUV Imager and Spectrograph for the ISS. AGU Abstract id.SH31C-08

Gonzalez Hernandez, I., Hill, F., Lindsey, C. (2007). Calibration of seismic signatures of active regions on the far side of the Sun. *Astrophys. J.* 669,1382–1389. doi:10.1086/521592

Gopalswamy, N., and Thompson, B. J. (2000). Early life of coronal mass ejections. *J. Atmos. Sol.-Terr. Phys.,* 62. 1457–1469, doi:10.1016/S1364-6826(00)00079-1

Gopalswamy, N. (2011). "Coronal Mass Ejections and Solar Radio Emissions," in Planetary Radio Emissions VII ed. H. O. Rucker, W. S. Kurth, P. Louarn, and G. Fischer (Vienna: Austrian Academy of Sciences Press), 325–342.

Gopalswamy, N., Davila, J. M., Auchère, F., Schou, J., Korendyke, C. M., Shih, A., et al., (2011a). Earth-Affecting Solar Causes Observatory (EASCO): A mission at the Sun–Earth L5, *Proc. SPIE 8148, Solar Physics and Space Weather Instrumentation IV*, 81480Z. doi:10.1117/12.901538

Gopalswamy, N., Davila, J. M., St. Cyr, O. J., Sittler, E. C., Auchère, F., Duvall, T. L., et al. (2011b). Earth-Affecting Solar Causes Observatory (EASCO): A potential international living with a star mission from Sun–Earth L5. *J. Atmos. Sol.-Terr. Phys.* 73, 658–663. doi:10.1016/j.jastp.2011.01.013

Gopalswamy, N., Yashiro, S., Akiyama, S., and Xie, H. (2018). Coronal Flux Ropes and Their Interplanetary Counterparts. *J. Atmos. Sol. Terr. Phys.* 180, 35–45. doi:10.1016/j.jastp.2017.06.004

Gopalswamy, N., Newmark, J., Yashiro, S., Mäkelä, P., Reginald, N., Thakur, N., et al. (2021). The Balloon-Borne Investigation of Temperature and Speed of Electrons in the Corona (BITSE): Mission Description and Preliminary Results. *Solar Phys.* 296, 15, doi:10.1007/s11207-020-01751-8

Gosain, S., Harvey, J., Martinez-Pillet, V., Woods, T. N., and Hill, F. (2022). A compact full-disk solar magnetograph based on miniaturization of GONG instrument. : *arXiv*,2207.07728. doi:10.48550/arXiv.2207.07728

Harrison, R. A., Davies, J. A., Rouillard, A. P., Davis, C. J., Eyles, C. J., Bewsher, D. et al. (2009). Two Years of the STEREO Heliospheric Imagers. Invited Review. *Solar Phys.* 256, 219, doi: 10.1007/s11207-009-9352-7

Harvey, J. W., Hill, F., Hubbard, R. P., Kennedy, J. R., Leibacher, J. W., Pintar, J. A., et al. (1996). The Global Oscillation Network Group (GONG) Project. *Science* 272, 1284–1286. doi:10.1126/science.272.5266.1284

Hassler, D. M., Gosain, S., Wuelser, J.-P., Harvey, J., Woods, T. N., Alexander, J., et al. (2022). The compact Doppler magnetograph (CDM) for solar polar missions and space weather research. *Proc. SPIE 12180, Space Telescopes and Instrumentation 2022: Optical, Infrared, and Millimeter Wave*, 121800K. doi:10.1117/12.2630663

Hassler, D. M., Newmark, J., Gibson, S., Harra, L., Appourchaux, T., Auchere, F., et al. (2020). The Solaris Solar Polar Mission. https://meetingorganizer.copernicus.org/EGU2020/EGU2020-17703.html [Accessed January 24, 2023].

Hill, F. (2018). The Global Oscillation Network Group Facility—An Example of Research to Operations in Space Weather. *Space Weather* 16, 1488–1497. doi:10.1029/2018SW002001

Howard, R. A., Moses, J. D., Vourlidas, A., Newmark, J. S., Socker, D. G., Plunkett, S. P. et al. (2008). Sun Earth Connection Coronal and Heliospheric Investigation (SECCHI). *Space Sci. Rev.* 136, 67–115. doi: 10.1007/s11214-008-9341-4

Howard, R. A., Vourlidas, A., Colaninno, R. C., Korendyke, C. M., Plunkett, S. P., Carter, M. T., et al. (2020). The Solar Orbiter Heliospheric Imager (SoloHI). *Astron. Astrophys.* 642, A13. doi:10.1051/0004-6361/201935202

Howard, T. A., Stovall, K., Dowell, J., Taylor, G. B., and White, S. M. (2016). Measuring the Magnetic Field of Coronal Mass Ejections Near the Sun Using Pulsars. *Astrophys. J.* 831, 208. doi:10.3847/0004-637X/831/2/208

Hurlburt, N. (2021). Imaging spectro-polarimeter using photonic integrated circuits. U.S. Patent No. 11099297B1. Washington, DC: U.S. Patent and Trade Office.

Hurlburt, N., Vasudevan, G. and Chintzoglou, G. (2022). IPSOS, the Imaging Photonic Magnetograph for Observing the Sun. https://sswrf.boulder.swri.edu/wordpress/wp-content/uploads/2022/10/SSWRF_abstracts.pdf [Accessed January 24, 2023].

Hurlburt, N., and the MICRO team (2023). Laboratory testing of a photonic magnetograph. *Astrophys. J.* (in prep.)

Hurlburt, N., and Berger, T. (2021). Architectures for Space Weather Magnetographs. https://www.hou.usra.edu/meetings/helio2050/pdf/4121.pdf [Accessed January 24, 2023].

Jensen, E. A. and Russell, C. T. (2008). Faraday rotation observations of CMEs. *Geophys. Res. Lett*. 35, CiteID L02103. doi: 10.1029/2007GL031038

Jensen, E. A., Gopalswamy, N., Wilson, L. B., III, Jian, L. K., Fung, S. F., Nieves-Chinchilla, T. et al. (2023). The Faraday Effect Tracker of Coronal and Heliospheric Structures (FETCH) instrument. *FrASS*. 10, id 1064069. 10.3389/fspas.2023.1064069

Jin, M., Manchester, W. B., van der Holst, B., Sokolov, I. V., Toth, G., Vourlidas, A. (2017a). Chromosphere to 1 AU Simulation of the 2011 March 7th Event: A Comprehensive Study of Coronal Mass Ejection Propagation. *Astrophys. J.* 834, 172. Doi: 10.3847/1538-4357/834/2/172

Jin, M., Manchester, W. B., van der Holst, B., Sokolov, I. V., Toth, G., Mullinix, R. E., et al. (2017b). Data-Constrained Coronal Mass Ejections in a Global Magnetohydrodynamics Model. *Astrophys. J.* 834, 173. doi: 10.3847/1538-4357/834/2/173

Kaiser, M. L., Kucera, T. A., Davila, J. M., St. Cyr, O. C., Guhathakurta, M., Christian, E. (2008). The STEREO Mission: An introduction. *Space Sci. Rev.* 136, 5–16. doi: 10.1007/s11214-007-9277-0

Kanekal, S. G., Blum, L., Christian, E. R., Crum, G., Desai, M., Dumonthier, J., et al. (2019). The MERiT onboard the CeREs: A novel instrument to study energetic particles in the Earth's radiation belts. *J. Geophys. Res. Space Phys.* 124, 5734–5760. doi:10.1029/2018JA026304

Kasper, J. C., Bale, S. D., Belcher, J. W., Berthomier, M., Case, A. W., Chandran, B. D. G., et al. (2019). Alfvénic velocity spikes and rotational flows in the near-Sun solar wind. *Nature* 576, 228–231. doi:10.1038/s41586-019-1813-z

Kooi, J. E., Wexler, D., Jensen, E. A., Kenny, M. N., Nieves-Chinchilla, T., Wilson III, L. B., et al. (2022). Modern Faraday Rotation Studies to Probe the Solar Wind. *Front. Astron. Space Sci*. 9, 841866. doi:10.3389/fspas.2022.841866

Koutchmy, S. (1988). Space-borne coronagraphy. *Space Sci. Rev.* 47, 95–143. doi:10.1007/BF00223238

Krucker, S., Hurford, G. J., Grimm, O., Kögl, S., Gröbelbauer, H.-P., Etesi, L., et al. (2020). The Spectrometer/Telescope for Imaging X-rays (STIX), *Astron. Astrophys.* 642, A15. doi: 10.1051/0004-6361/201937362

Krupar, V., Santolik, O., Cecconi, B., Maksimovic, M., Bonnin, X., et al. (2012). Goniopolarimetric inversion using SVD: An application to type III radio bursts observed by STEREO. *J. Geophys. Res*., 117, A06101. doi:10.1029/2011JA017333

Lavraud, B., Liu, Y., and Segura, K. (2016). A small mission concept to the Sun-Earth Lagrangian L5 point for innovative solar, heliospheric and space weather science. *J. Atmos. Solar Terr. Phys.* 146, 171–185. doi:10.1016/j.jastp.2016.06.004

Lemen, J. R., Title, A. M., Akin, D. J., Boerner, P. F., Chou, C., Drake, J. F., et al. (2012). The Atmospheric Imaging Assembly (AIA) on the Solar Dynamics Observatory (SDO**)**. *Solar Phys*. 275, 17–40. doi:10.1007/s11207-011-9776-8

Liu, Y., Manchester, W. B., IV, Kasper, J. C., Richardson, J. D., and Belcher, J. W. (2007). Determining the Magnetic Field Orientation of Coronal Mass Ejections from Faraday Rotation. *Astrophys. J.* 665, 1439. doi:10.1086/520038

Mäkelä, P., Gopalswamy, N., Reiner, M. J., Akiyama, S., and Krupar, V. (2016). Source Regions of the Type II Radio Burst Observed During a CME-CME Interaction on 2013 May 22. *Astrophys. J.* 827, 141. doi:10.3847/0004-637X/827/2/141

Manchester, W. B., van der Holst, B., Lavraud, B. (2014). Flux rope evolution in interplanetary coronal mass ejections: the 13 May 2005 event. Plasma Phys. Control Fusion 56, 1-11. Doi: 10.1088/0741-3335/56/6/064006

Mancuso, S., and Garzelli, M. V. (2013). Coronal magnetic field strength from Type II radio emission: complementarity with Faraday rotation measurements. *Astron. Astrophys.* 460, L1. doi:10.1051/0004-6361/201322645

Masuda, S., Kosugi, T., Hara, H., Tsuneta, S., and Ogawara, Y. (1994). A loop-top hard X-ray source in a compact solar flare as evidence for magnetic reconnection. *Nature* 371, 495–497. doi:10.1038/371495a0

Mays, M. L., Thompson, B. J., Jian, L. K., Colaninno, R. C., Odstrcil, D., Möstl, C., et al. (2015). Propagation of the 7 January 2014 CME and Resulting Geomagnetic Non-event. *Astrophys. J.* 812, 145. doi:10.1088/0004-637X/812/2/145

Müller, D., St. Cyr, O. C., Zouganelis, I., Gilbert, H. R., Marsden, R., Nieves-Chinchilla, T. et al. (2020). The Solar Orbiter mission. Science overview. Astronomy & Astrophysics, 642, id.A1, doi:10.1051/0004-6361/202038467

Na, H., Moon, Y. -J., and Lee, H. (2017). Development of a Full Ice-cream Cone Model for Halo Coronal Mass Ejections. *Astrophys. J.* 839, 82. doi:10.3847/1538-4357/aa697c

Na, H., Moon, Y. -J., Jang, S., Lee, K.-S., and Kim, H.-Y. (2013). Comparison of Cone Model Parameters for Halo Coronal Mass Ejections. *Solar Phys.* 288, 313–329. doi:10.1007/s11207-013-0293-9

Odstrcil, D. (2003). Modeling 3-D solar wind structure. *Adv. Space Res.* 32, 497. doi:10.1016/S0273-1177(03)00332-6

Odstrcil, D., Mays, M. L., Hess, P., Jones, S. I., Henney, C. J., and Arge, C. N. (2020). Operational Modeling of Heliospheric Space Weather for the Parker Solar Probe. *Astrophys. J. Suppl. Ser.*, 246, 73. doi:10.3847/1538-4365/ab77cb

Petrie, G., Pevtsov, A., Schwarz, A., and DeRosa, M. (2018). Modeling the Global Coronal Field with Simulated Synoptic Magnetograms from Earth and the Lagrange Points L3, L4, and L5. *Solar Phys.* 293, 88. doi:10.1007/s11207-018-1306-5

Pevtsov, A. A.. Petrie, G.. MacNeice, P., and Virtanen, I. I. (2020). Effect of Additional Magnetograph Observations From Different Lagrangian Points in Sun-Earth System on Predicted Properties of Quasi-Steady Solar Wind at 1 AU. *Space Weather* 18, e02448. doi:10.1029/2020SW002448

Pesnell, W. D., Thompson, B. J., and Chamberlin, P. C. (2012). The Solar Dynamics Observatory. *Solar Phys.* 275, 3–15. doi: 10.1007/s11207-011-9841-3Pomoell, J., and Poedts, S. (2018). EUHFORIA: European heliospheric forecasting information asset. *J. Space Weather Space Clim.* 8, A35. doi:10.1051/swsc/2018020

Posner, A., Arge, C. N., Staub, J., StCyr, O. C., Folta, D., Solanki, S. K. Et al. (2021). A Multi-Purpose Heliophysics L4 Mission. Space Weather 19, article id. e02777, doi: 10.1029/2021SW002777

Reinisch, B.W., D.M. Haines, K. Bibl, G. Cheney, I.A. Galkin, X. Huang, S.H.et al. (2000), The Radio Plasma Imager investigation on the IMAGE spacecraft. *Space Sci. Rev.*, 91, 319-359, doi:10.1023/A:1005252602159.

Sachdeva, N., van der Holst, B., Manchester, W. B. IV, Tóth, G., Chen, Y., Lloveras, D. G. (2019). Validation of the Alfvén Wave Solar Atmosphere Model (AWSoM) with Observations from the Low Corona to 1 au. *Astrophys. J.* 887, 63, doi: 10.3847/1538-4357/ab4f5e

Sachdeva, N., Tóth, G., Manchester, W. B., van der Holst, B., Huang, Z., et al. (2021). Simulating Solar Maximum Conditions Using the Alfvén Wave Solar Atmosphere Model (AWSoM). *Astrophys. J.* 923, 176, doi: 10.3847/1538-4357/ac307c

Scherrer, P. H., Bogart, R. S., Bush, R. I., Hoeksema, J. T., Kosovichev, A. G., et al. (1995). The Solar Oscillations Investigation - Michelson Doppler Images (MDI). *Solar Phys.* 162, 129. doi:10.1007/BF00733429

Scherrer, P. H., Schou, J., Bush, R. I., Kosovichev, A. G., Bogart, R. S. et al. (2012). The Helioseismic and Magnetic Imager (HMI) Investigation for the Solar Dynamics Observatory (SDO). *Solar Phys.* 275, 207. doi:10.1007/s11207-011-9834-2

Scolini, C., Rodriguez, L., Mierla, M., Pomoell, J., and Poedts, S. (2019). Observation-based modelling of magnetised coronal mass ejections with EUHFORIA. *Astron. Astrophys.*, 626, A122. doi:10.1051/0004-6361/20193505

Scolini, C., Chané, E., Temmer, M., Kilpua, E. K. J., Dissauer, K., Veronig, A. M., et al. (2020). CME-CME Interactions as Sources of CME Geoeffectiveness: The Formation of the Complex Ejecta and Intense Geomagnetic Storm in 2017 Early September. *Astrophys. J. Suppl. Ser.* 247, 21. doi:10.3847/1538-4365/ab6216

Seaton, D. B., and Darnel, J. M. (2018). Observations of an Eruptive Solar Flare in the Extended EUV Solar Corona. *Astrophys. J. Lett.* 852, L9. doi:10.3847/2041-8213/aaa28e

Shiota, D., and Kataoka, R. (2016). Magnetohydrodynamic simulation of interplanetary propagation of multiple coronal mass ejections with internal magnetic flux rope (SUSANOO-CME). *Space Weather* 14, 56–75. doi:10.1002/2015SW001308

Singh, T., Yalim, M. S., Pogorelov, N. V., and Gopalswamy, N. (2019). Simulating Solar Coronal Mass Ejections Constrained by Observations of Their Speed and Poloidal Flux, *Astrophys. J. Lett.* 875, L17, doi:10.3847/2041-8213/ab14e9

Socker, D. G., Howard, R. A., Korendyke, C. M., Simnett, G. M., and Webb, D. (2000). NASA Solar Terrestrial Relations Observatory (STEREO) mission heliospheric imager. *Proc. SPIE 4139, Instrumentation for UV/EUV Astronomy and Solar Missions.* doi:10.1117/12.410528

Solanki, S. K., del Toro Iniesta, J. C., Woch, J., Gandorfer, A., Hirzberger, J., Alvarez-Herrero, A. et al. (2020). The Polarimetric and Helioseismic Imager on Solar Orbiter. *Astron. Astrophys.* 642, A11. Doi: 10.1051/0004-6361/201935325

Staub, J., Fernandez-Rico, G., Gandorfer, A., Gizon, L., Hirzberger, J., Kraft, S. et al. (2020). PMI: The Photospheric Magnetic Field Imager. *J. Space Weather Space Clim.* 10, id.54. doi: 10.1051/swsc/2020059

Temmer, M. (2021). Space weather: the solar perspective. *Living Rev Sol Phys* 18, 4, doi: 10.1007/s41116-021-00030-3

Thernisien, A., Howard, R. A. and Vourlidas, A. (2006). Modeling of Flux Rope Coronal Mass Ejections. *Astrophys. J.* 652, 763–773. doi:10.1086/508254

Thernisien, A., Vourlidas, A., and Howard, R. A. (2009). Forward Modeling of Coronal Mass Ejections Using STEREO/SECCHI Data, *Solar Phys.* 256, 111–130. doi:10.1007/s11207-009-9346-5

Török, T., Downs, C., Linker, J. A., Lionello, R., Titov, V. S., Mikić, S., et al. (2018). Sun-to-Earth MHD Simulation of the 2000 July 14 "Bastille Day" Eruption. *Astrophys. J.* 856, 75, doi:10.3847/1538-4357/aab36d

Van der Holst, B., Sokolov I. V., Meng, X., Jin, M., Manchester, W. B., Tóth, G., et al. (2014). Alfvén Wave Solar Model (AWSOM): Coronal Heating. *Astrophys. J.* 782, 81, doi: 10.1088/0004-637X/782/2/81

Vourlidas, A., Howard, R. A., and Plunkett, S. P. (2016). The Wide-Field Imager for Solar Probe Plus (WISPR). *Space Sci. Rev.* 204, 83–130. doi:10.1007/s11214-014-0114-y

Verbeke, C., Mays, M. L., Kay, C., Riley, P., Palmerio, E., Dumbović, M., et al. (2022). Quantifying errors in 3D CME parameters derived from synthetic data using white-light reconstruction techniques. *Adv. Space Res.* (in press). doi:10.1016/j.asr.2022.08.056

Wexler, D., Manchester, W.B., Lian, J. K., Wilson III, L. B., Gopalswamy, N., Song, P. et al. (2023). Investigating A Solar Wind Stream Interaction Region using Interplanetary Spacecraft Radio Signals: An MHD simulation Study. *Astrophys. J.* (in press)

Wuelser, J., Lemen, J. R., Nitta, N. V. (2007). The STEREO SECCHI/EUVI EUV coronal imager. In Solar Physics and Space Weather Instrumentation II. Edited by S. Fineschi and R. A. Viereck. *Proceedings of the SPIE*, Vol. 6689, article id. 668905. doi: 10.1117/12.747563

Yang, D., Gizon, L., Barucq, H., Hirzberger, J., Orozco Suárez, D., Albert, K. et al. (2023). Direct assessment of SDO/HMI helioseismology of active regions on the Sun's far side using SO/PHI magnetograms. *A&A* 674, A183, doi:10.1051/0004-6361/202346030

Zhao, X. P., Plunkett, S. P., and Liu, W. (2002). Determination of geometrical and kinematical properties of halo coronal mass ejections using the cone model. *J. Geophys. Res. Space Phys.* 107, 1223. doi:1 0.1029/2001JA009143